\documentclass{aa}				
\usepackage{graphicx}
\usepackage{txfonts}

\usepackage{natbib,twoopt}

\bibpunct{(}{)}{;}{a}{}{,} % to follow the A&A style

\usepackage[breaklinks=true]{hyperref} %% to avoid \citeads line fills
\bibpunct{(}{)}{;}{a}{}{,} %% natbib format for A&A and ApJ
\makeatletter
\newcommandtwoopt{\citeads}[3][][]{\href{http://adsabs.harvard.edu/abs/#3}%
{\def\hyper@linkstart##1##2{}%
\let\hyper@linkend\@empty\citealp[#1][#2]{#3}}}
\newcommandtwoopt{\citepads}[3][][]{\href{http://adsabs.harvard.edu/abs/#3}%
{\def\hyper@linkstart##1##2{}%
\let\hyper@linkend\@empty\citep[#1][#2]{#3}}}
\newcommandtwoopt{\citetads}[3][][]{\href{http://adsabs.harvard.edu/abs/#3}%
{\def\hyper@linkstart##1##2{}%
\let\hyper@linkend\@empty\citet[#1][#2]{#3}}}
\newcommandtwoopt{\citeyearads}[3][][]%
{\href{http://adsabs.harvard.edu/abs/#3}
{\def\hyper@linkstart##1##2{}%
\let\hyper@linkend\@empty\citeyear[#1][#2]{#3}}}
\makeatother

\begin{document}

\title{Calibrating and Stabilizing Spectropolarimeters with Charge Shuffling and Daytime Sky Measurements}

\titlerunning{Calibrating and stabilizing spectropolarimeters}

\author{David Harrington \inst{1} \inst{2} \inst{3} \and J.R. Kuhn \inst{4} \and Rebecca Nevin \inst{5} }
\authorrunning{Harrington et. al.}

\institute{Kiepenheuer-Institut f\"{u}r Sonnenphysik, Sch\"{o}neckstr. 6, D-79104 Freiburg, Germany \\
\and Institute for Astronomy, University of Hawaii, 2680 Woodlawn Drive, Honolulu, HI, 96822, USA  \\
\and Applied Research Labs, University of Hawaii, 2800 Woodlawn Drive, Honolulu, HI, 96822, USA \\
\and  Institute for Astronomy Maui, University of Hawaii, 34 Ohia Ku St., Pukalani, HI, 96768, USA \\
\and  Department of Astrophysical and Planetary Sciences, University of Colorado, Boulder, CO 80309, USA  }

\date{Submitted October, 2013}

\abstract{Well-calibrated spectropolarimetry studies at resolutions of $R>$10,000 with signal-to-noise ratios (SNRs) better than 0.01\% across individual line profiles, are becoming common with larger aperture telescopes. Spectropolarimetric studies require high SNR observations and are often limited by instrument systematic errors. As an example, fiber-fed spectropolarimeters combined with advanced line-combination algorithms can reach statistical error limits of 0.001\% in measurements of spectral line profiles referenced to the continuum. Calibration of such observations is often required both for cross-talk and for continuum polarization. This is not straightforward since telescope cross-talk errors are rarely less than $\sim$1\%. In solar instruments like the Daniel K. Inouye Solar Telescope (DKIST), much more stringent calibration is required and the telescope optical design contains substantial intrinsic polarization artifacts. This paper describes some generally useful techniques we have applied to the HiVIS spectropolarimeter at the 3.7m AEOS telescope on Haleakala. HiVIS now yields accurate polarized spectral line profiles that are shot-noise limited to 0.01\% SNR levels at our full spectral resolution of 10,000 at spectral sampling of $\sim$100,000. We show line profiles with absolute spectropolarimetric calibration for cross-talk and continuum polarization in a system with polarization cross-talk levels of essentially 100\%. In these data the continuum polarization can be recovered to one percent accuracy because of synchronized charge-shuffling model now working with our CCD detector. These techniques can be applied to other spectropolarimeters on other telescopes for both night and day-time applications such as DKIST, TMT and ELT which have folded non-axially symmetric foci.}


\maketitle

\section{Introduction}

Polarimetry and spectropolarimetry generally require high signal-to-noise observations from intrinsically imperfect telescopes and instruments. The fundamentally differential nature of polarimetric measurements is an advantage for minimizing systematic instrumental noise. Nevertheless, achieving photon-noise limited accuracy is often dependent on mitigating subtle instrument systematics. These errors fall into two categories: I) calibration uncertainty, and II) measurement instability. We describe here a strategy for minimizing both of these uncertainties using new technologies and algorithms in the context of stellar spectropolarimetry.

Measured optical polarization is sensitive to the local geometry of the radiating or scattering source and allows otherwise optically unresolved features of the source to be inferred using forward modeling approaches.  We have been investigating polarimetric signatures in individual spectral lines of young stars that show Stokes QU features with amplitudes of about 0.01\% \citep{Harrington:2010km, Harrington:2009gy, Harrington:2008jq}. Such tiny polarized spectral features can contain basic information about, for example, the geometry of the circumstellar environment \citep{Kuhn:2011tk,Kuhn:2007fv}. 
 
This problem of measuring small polarized features in astronomical objects is not at all unique to stellar spectropolarimetry and has a long heritage beginning in solar magnetic field measurements. Most measurement techniques that deliver such high precision usually follow a multi-step process. Some type of modulation occurs where polarization signals are optically converted to intensity variations. To de-modulate these variations, several independent exposures are somehow differenced to remove some systematic errors. These differential measurements then have to be calibrated to ultimately derive an absolute polarization of the input signal. Several calibration techniques are usually required to fully recover the input polarization state of the source from the polarizing influence of the modulators, telescope and detectors. One important calibration need is to correct the scrambling of the input polarization state from the observed output polarization state (called cross-talk), that is caused by the telescope and detectors. At a lower level telescopes also often create or destroy the degree of polarization of the input light. This induced polarization and depolarization, must also be calibrated. Different modulation strategies, beam differencing techniques and technical considerations drive instruments to very different approaches depending on the science goals and practical limitations. For some specific information on typical modulation schemes and efficiencies, see \citet{Keller:2009vj, Tomczyk:2010wta, Wijn:2011wt}, astronomical measurement techniques and specific stellar applications, see \citet{2013pss2.book..175S}.

Calibrating night-time spectropolarimetry generally requires data from both known polarized and unpolarized sources. Other techniques make assumptions about for example, the polarized spectral symmetry of the source; or partial knowledge about the polarizing properties of the telescope and instrument. For example, solar observations can be calibrated to a high level of accuracy using observations of sources with a fundamental wavelength polarization symmetry \citep{Kuhn:1994jk}. In all cases the full Stokes spectra must be measured in order to achieve the most precise absolute polarization calibration. Under certain limitations and assumptions, the number of measurements can be reduced for efficiency reasons. For example, instruments with relatively low cross talk can neglect to record some parts of the Stokes vector if they are deemed unnecessary or inefficient. 

Minimizing type II ("stability") polarization uncertainty relies mostly on fixing instrument, telescope, and atmosphere variability during the modulation process where differencing intensity measurements is necessary to derive the full-Stokes polarized spectra. Techniques like dual beam analyzers, fast-modulation, beam swapping modulation, or other instrument stabilizing schemes decrease this noise in spectropolarimeter+telescope measurements.  Where many photons are needed, obtaining efficient calibrations is especially important. The major limitations are often observing duty-cycle, instrumental drifts, waste of precious night time to calibrate, or an inability to calibrate at all. We demonstrate here a new instrument using a combination of techniques that helps minimize several of these error sources at a highly non-optimal polarimetric telescope focus. With these techniques we can deliver both stability and high differential accuracy with reasonable absolute polarization calibration without rebuilding the telescope to be optimized for polarimetry.

\subsection{Spectropolarimetric instruments, errors \& suppression techniques}

	Large aperture telescopes are tasked for various performance requirements. For example, some are optimized to deliver accurate continuum polarization, low cross-talk, or high spectral resolution and sensitivity. On short timescales telescope jitter, tracking errors and atmospheric seeing can change the footprint of a beam on the telescope and spectrograph optics and can modulate the detected intensity to create spurious polarization. On long timescales optical coatings oxidize \citep{vanHarten:2009gi} making calibrations unstable in time. As non-equatorial telescopes track a target, mirrors rotate with respect to the field to change the properties of the optical system. Scattered light and instrument instabilities in a spectrograph can cause drifts. The electronic properties of the detector pixels, readout amplifiers and signal processing algorithms imprint flat-fielding errors, bias drifts and other non-linearities into the data. Complex optical paths such as coud\'{e} foci involve fold mirrors at oblique angles that induce strong polarization effects.
	
	Many of these errors can be overcome by using calibration techniques or by designing special-purpose instruments. Formal polarimetric error budgets applied over wide ranges of science cases are typically not implemented but formalisms to do so are becoming more common \citep{Keller:2009vj, ovelar:2012ov, ovelar:2012elt}. Dual-beam polarimeters record two orthogonally analyzed beams in a single exposure and double the instrument efficiency. Most instruments use beam-exchange or other modulation techniques to difference subsequent exposures and remove many effects caused by detector electronics, flat-fielding and optics behind the analyzer.  
	
	In almost all dual-beam night-time spectropolarimeters, several independent exposures are required to make a full-Stokes measurement. The timescale for these exposures (and polarimetric modulation) is often many minutes. For instance, there are several existing and planned fiber-fed instruments like ESPaDOnS, HARPSpol, BOES, PEPSI that are designed to deliver spectral resolutions above 50,000 and large one-shot wavelength coverage by using cross-dispersion \citep{Semel:1993vy, Donati:1997wj, Donati:1999dh, 2011Msngr.143....7P, 2003SPIE.4843..425M, Snik:2011vm, Kim:2007ja, Snik:2008fh,Strassmeier:2008ho}. These instruments are not designed to measure continuum polarization but fairly reasonable measurement precision is possible under a range of circumstances \footnote{\tiny{http://www.cfht.hawaii.edu/Instruments/Spectroscopy/Espadons/ContiPolar/}} \citep{Pereyra:2015gt}. However, these instruments do not have spatial resolution due to the fiber-feed. In addition, absolute calibration can be limited by cross-talk systematic errors in up-stream optics even at Cassegrain focus. For ESPaDOnS, several years of observations revealed time-dependent cross-talk errors from its atmospheric dispersion compensator and collimating lens. These effects limited absolute calibrations to the $>$3\% level in the initial few years of data. Improved optics and testing reduced the cross-talk to below the 1\% design thresholds, but it is still present as the major calibration limitation \citep{Barrick:2010jz,Barrick:2010gv, Pereyra:2015gt}

	Other night-time slit-fed instruments have different performance capabilities and correspondingly different stability issues such as FORS on the VLT \citep{Bagnulo:2012dr,Seifert:2000vs,Bagnulo:2002cd,Bagnulo:2009bz,Patat:2006en}.  Cassegrain instruments such as LRISp on Keck, ISIS on the WHT are examples of variable-gravity spectropolarimeters which are designed to deliver relatively stable continuum polarization, imaging capabilities and more modest resolutions but they can be subject to mechanical flexure or environment "drifts" unless designed and stabilized properly \citep{Goodrich:2003kv, Goodrich:1995fg, Leone:2007gs, Bjorkman:2005tx, Lomax:2012eo, Nordsieck:2001wn, 1996ASPC...97..100N, Eversberg:1998fh}.

	Its often true that cross-talk errors are strong and time-dependent in instruments where polarimetry was not a driving requirement. Many large night-time telescope projects (such as the TMT, GMT or ELT) do not allow access to Cassegrain or Gregorian foci for traditional polarimetric instruments with low cross-talk. Some of the most accurate single line spectropolarimetric measurements ever obtained have been realized by off-axis telescopes \citep{Lin:2004hq}. The great advantages for dynamic range and scattered light stability of such telescopes outweighs the fractional percent polarization induced by the asymmetric primary optics. For instance, the Daniel K Inouye Solar Telescope (DKIST) will make good use of several high-resolution spectropolarimeters operating at the $10^{-4}$ precision levels. This system relies on several precise calibration techniques that impart stringent requirements on the telescope Mueller Matrix \citep{ Keil:2011wj, Keller:2003uj, Nelson:2010fw, deWijn:2012dd, SocasNavarro:2005bq}. As projects grow in scope and complexity, real error analysis and tradeoff studies are becoming essential as exemplified by the VLT and E-ELT codes under development \citep{ovelar:2012ov, ovelar:2012elt}. Even axially symmetric telescope beams require calibration of instrumental polarization from the instrument optics and non-uniformities in coatings \citep{Hough:2006iz, Bailey:2010de, Bailey:2008fm, Wiktorowicz:2008fm}

	In the solar community, several instruments have included very fast polarimetric modulation to overcome many of these common instrument issues to achieve 0.001\% differential precision polarimetry.  Modulators include liquid crystals (LCs), piezo-elastic modulators (PEMs), rapidly rotating components and other techniques \citep{Gandorfer:1999fq, 1992SPIE.1746...22E, Skumanich:1997eh, Gandorfer:2004fm, Lin:2004hq}. Modulation can be chromatically balanced, tuned or optimized for various observing cases \citep{Povel:1995dn, Gisler:2003hy, Tomczyk:2010wta, deWijn:2010fh, Wijn:2011wt, Snik:2009va, LopezAriste:2011wc, Nagaraju:2007tn}. At modulation rates faster than 1kHz most instrument effects and even atmospheric seeing fluctuations can be suppressed \citep{Keller:1994ww, Stenflo:2007wq, Hough:2006iz, Hanaoka:2004ku, Rodenhuis:2012du, Xu:2006hk}. Night-time modulation frequencies are usually much slower and fixed at the exposure time so that detector read noise does not dominate. 

	Other strategies to calibrate, stabilize and optimize spectropolarimeters with varying degrees of cross-talk have been described (cf. \cite{Elmore:2010ip, Snik:2006iw, Tinbergen:2007fd, SocasNavarro:2011gn, delToroIniesta:2000cg, Tinbergen:2007fd, Witzel:2011ga, Bagnulo:2009bz, Bagnulo:2012dr, Giro:2003il, Witzel:2011ga, Maund:2008go}). Some use spectral modulation or other techniques to accomplish instrument stabilization \citep{Snik:2009va}. Other polarimeter designs, such as the VLT X-Shooter have controlled chromatic properties and use special techniques to calibrate \citep{Snik:2012jw}. 

	In night-time spectropolarimeters, magnetic field studies of stars are pushing precision limits by co-adding many lines into a single {\it pseudospectral line} that yields effective signal-to-noise ratios above 100,000. Large surveys such as MIMES \footnote{http://www4.cadc-ccda.hia-iha.nrc-cnrc.gc.ca/MiMeS/} and others use ESPaDOnS and NARVAL fiber-fed spectropolarimeters at spectral resolutions greater than 60,000 are examples \citep{Wade:2010vv, 2012AIPC.1429...67G, 2012ASPC..464..405W}. These programs combine many individual spectral lines over many spectral orders to derive a single representative {\it pseudo line profile} \citep{Donati:1997wj, Kochukhov:2010ig, Sennhauser:2009ds, 2013A&A...554A..93K, Petit:2011cu}. 

\begin{figure*} [!h, !t, !b]
\begin{center}
\includegraphics[width=0.99\linewidth, angle=0]{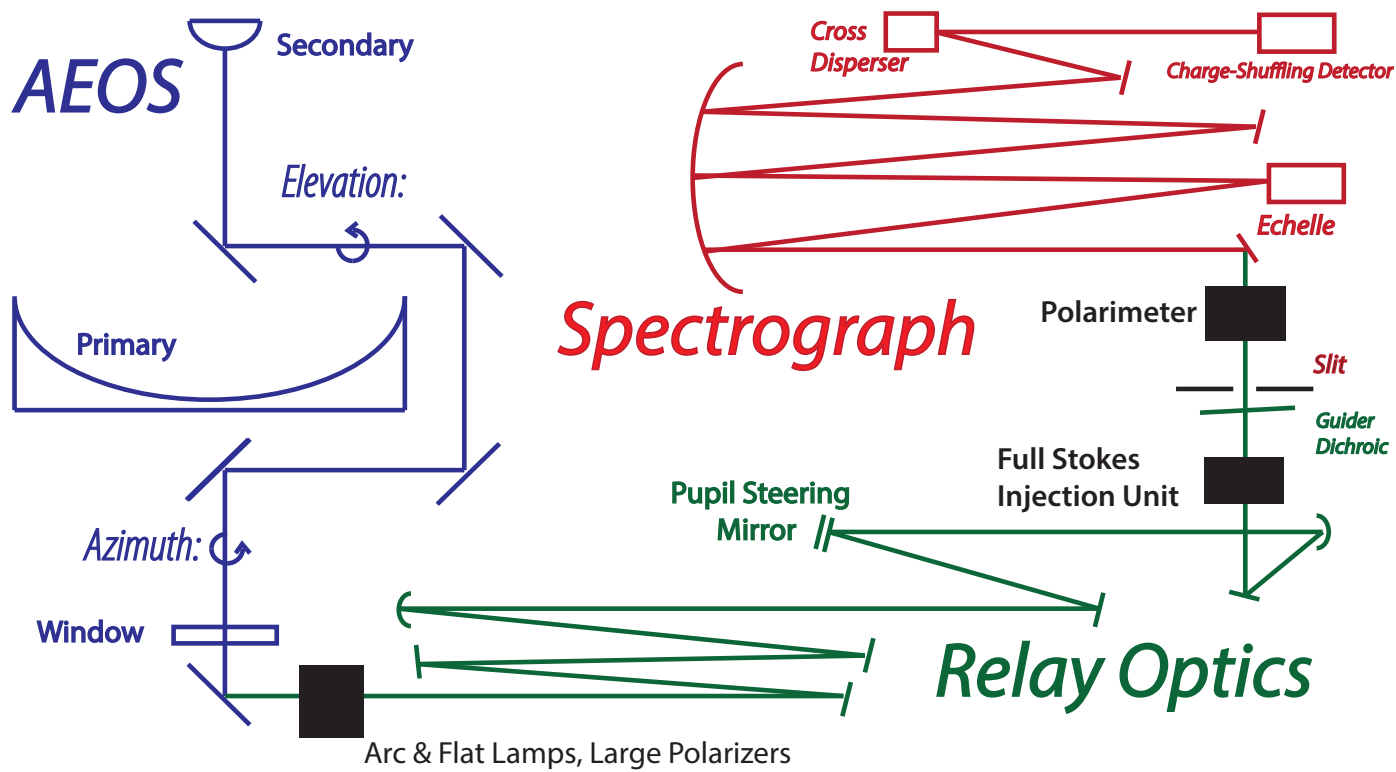}
\caption{ \label{hivis_block} This Figure shows a schematic layout for the AEOS and HiVIS optical pathway when in the charge-shuffling configuration. The telescope optics are shown in blue. There are 5 flat fold mirrors that bring the beam in to the coud\'{e} room. The elevation and azimuth axes are labeled to show the optical changes during telescope tracking. The HiVIS optical relay which feeds both visible and infrared spectrographs and accomplishes tip/tilt correction is shown in green. The spectrograph optics are shown in red. All optical surfaces are shown. Polarization and calibration components are shown in black. The HiVIS modulator and analyzer are behind the spectrograph slit. All optics in blue and green contribute to telescope polarization cross-talk. The Full-Stokes Injection Unit in front of the HiVIS spectrograph slit is for polarization calibration of the modulators and spectrograph.  Additional calibration optics shown in black between the AEOS telescope (blue) and relay optics (green) are mounted at the entrance of the coud\'{e} room. These provide flat fielding, wavelength and polarization calibration light sources. See text for details.}
\end{center}
\end{figure*}

	In other night-time science applications, single-line studies at more modest resolutions of 3,000 to 10,000 are sufficient but require similarly high signal-to-noise ratios across individual lines. For instance, Herbig Ae/Be stars and other emission-line stars show polarization signatures at the 0.1\% level and below. However, individual lines have varying profiles and line-combination techniques don't apply and sufficient SNRs are required to see individual line profiles \citep{Harrington:2009gy, Harrington:2009dz, Harrington:2008jq,Harrington:2007bq, Harrington:2011uj}.  Herbig Ae/Be and T-Tauri polarized line profiles can be quite large and variable when observed with effective resolutions of even 3,000 \citep{Vink:2005cp,Vink:2005ft,Vink:2002hf}. Absolute continuum polarization is a particularly important measurement for Be star and some other emission-line studies because interpreting the polarizing mechanisms depends on knowing polarization direction to distinguish models for the source region \citep{Bjorkman:2006te,Bjorkman:2006ts,Bjorkman:2000vn,Bjorkman:2000vc,Brown:1977vq,Carciofi:2009bu,Harries:1997ca,Harries:1996uw,Ignace:2000tl,Ignace:1999bl,Oudmaijer:1999hl,Oudmaijer:2005ik,Quirrenbach:1997hm}.  Wolf-Rayet stars can be similarly probed \citep{Harries:2000wk, Vink:2010vt}. Its important to note that there are a large number of interesting stellar problems where spectropolarimetric resolutions as low as 3000 are sufficient to learn about the source region. 

	We upgraded AEOS HiVIS spectropolarimeter to minimize several sources of instrumental instabilities to achieve high SNRs and reasonable calibration accuracies by using liquid crystal modulation and charge-shuffling detectors. HiVIS has a spectral sampling of 4.5pm to 5.5pm per pixel corresponding to a 1-pixel sampling of R$\sim$150,000. For this campaign we use a wide slit delivering a spectral resolution of 12,000 which is over-sampled by a factor of about 9 when charge shuffling. SNRs across the H$_\alpha$ line profile of 0.01\% are now achievable in this mode on fairly bright stars in a single night while delivering accurate continuum polarization with a stable polarization cross-talk calibration. We have implemented faster modulation techniques combined with electronic charge-shuffling to achieve this instrument stability. We have also improved calibration techniques using the daytime sky, demodulation techniques and other polarimetric optics in order to achieve reasonable continuum polarization stability and absolute polarization calibration.

	We show here that night-time long-slit spectropolarimetric observations can be calibrated for cross-talk to an absolute polarization reference frame to roughly $\sim$1\% levels in the presence of 100\% cross-talk. The polarization calibration changes with wavelength are generally smooth functions that can be measured at much lower spectral resolution. When calibrating spectral line profile measurements with 0.01\% signal-to-noise limits, these absolute calibrations can be treated simply as constants. Other technologies we use allow us to stabilize the stellar continuum polarization to a 0.1\% calibration level. These combined techniques allow us to deliver both high differential precision across line profiles while maintaining good absolute polarization calibration even in the presence of several difficult instrumental challenges.

\subsection{HiVIS and the charge-shuffling spectropolarimeter}

	We have been upgrading the High Resolution Visible and Infra-red Spectrograph (HiVIS) on the 3.67m Advanced Electro-Optical System (AEOS) telescope on Haleakala, Maui \citep{Harrington:2010km,Harrington:2008jq,Harrington:2006hu, 2008PhDT.........9H, Harrington:2011fz,Thornton:2003bi}. HiVIS is a long-slit spectrograph built for spectral resolutions of 10,000 to 50,000 in the AEOS coud\'{e} room. HiVIS is a cross-dispersed echelle spectrograph designed to deliver complete wavelength coverage with high order overlap across 15-20 spectral orders on a 4k by 4k focal plane mosaic. An optical layout of the spectrograph and polarimeter can be found in \citet{Harrington:2006hu}, \citet{Thornton:2003bi} and \citet{2008PhDT.........9H}. Figure \ref{hivis_block} shows a schematic layout of the telescope and spectropolarimeter.  There are 8 optical elements before the beam is delivered to the coud\'{e} room. The primary and secondary mirror create a converging f/200 beam after the secondary mirror reflection. This beam is delivered to the coud\'{e} room by 5 fold mirrors at 45$^\circ$ reflection angles. There is also a selectable window of either BK7 or Infrasil between the last two fold mirrors.  After entering the coud\'{e} room, there are 8 more reflections before the slit and the polarimeter in this HiVIS charge-shuffling configuration. These optics accomplish collimation, tip/tilt steering at a pupil image and basic optical packaging. With this complex optical layout, our instrument causes linear-to-circular cross talk of 100\% at some telescope pointings and wavelengths. At the entrance to the coud\'{e} room we have a calibration unit with arc, and flat field lamps..  There is also a polarization calibration unit consisting of an achromatic quarter wave plate and a wire grid polarizer under computer controlled rotation. The AEOS telescope has the largest telescope polarization we have experienced and is a challenging system to calibrate.

	We describe here the charge shuffling and liquid-crystal modulator improvements and the calibration algorithms we use for obtaining accurate full-stokes spectropolarimetry of stellar and circumstellar, visible and near-IR sources. These techniques yield continuum polarization stable to $<$0.1\% accuracy across all spectral orders.

\begin{figure*} [!h, !t, !b]
\begin{center}
\includegraphics[width=0.32\linewidth, angle=90]{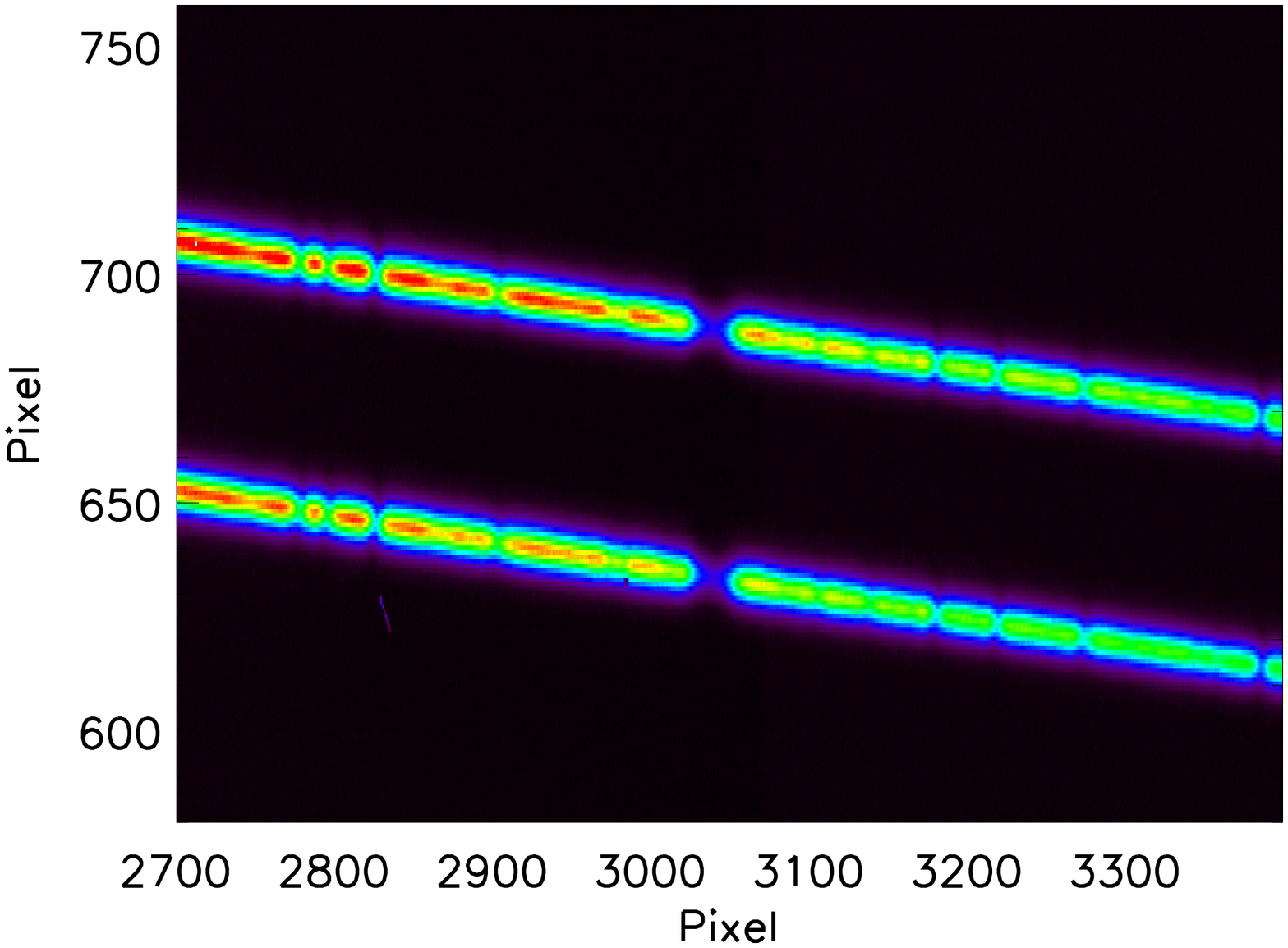}
\includegraphics[width=0.32\linewidth, angle=90]{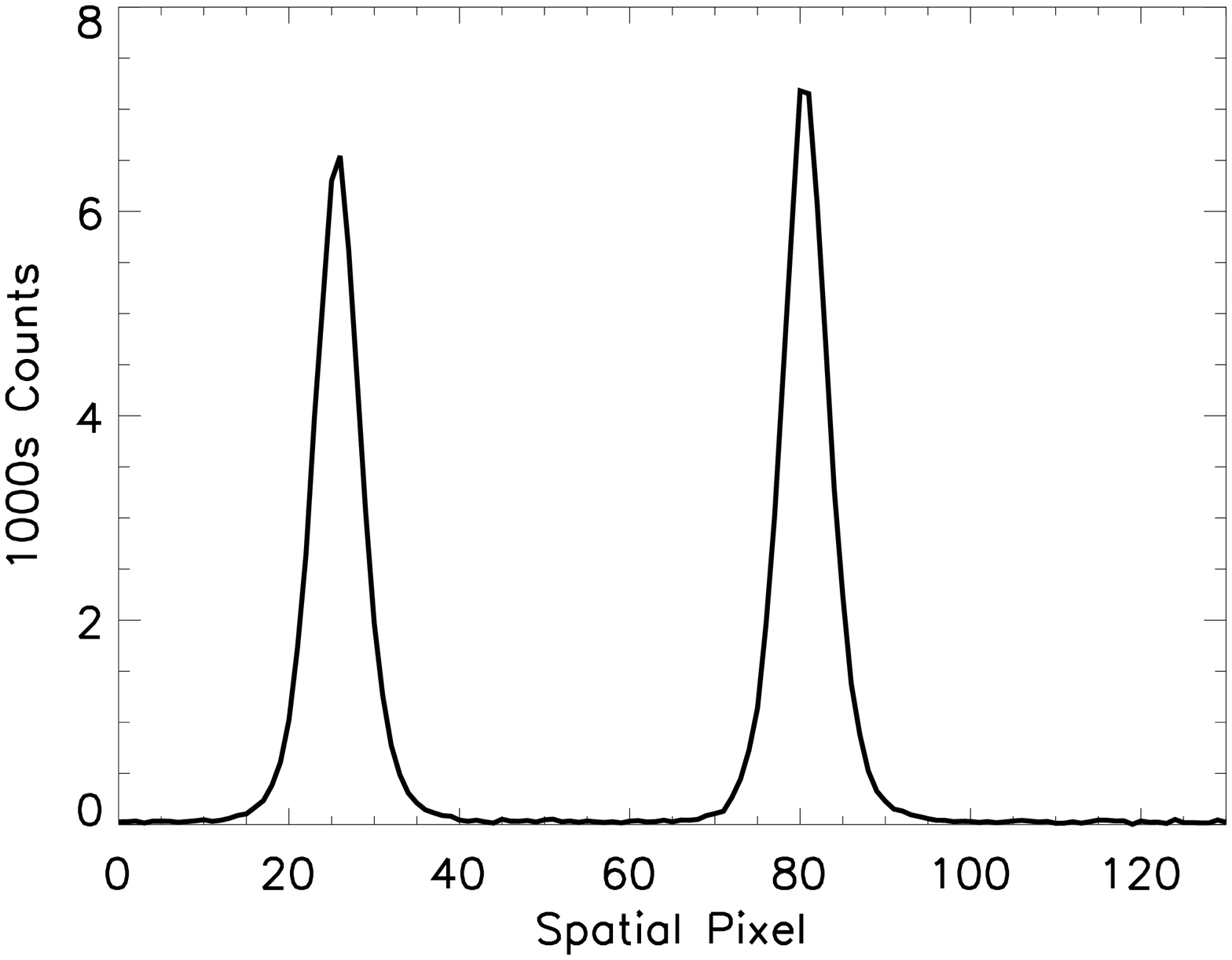}
\includegraphics[width=0.32\linewidth, angle=90]{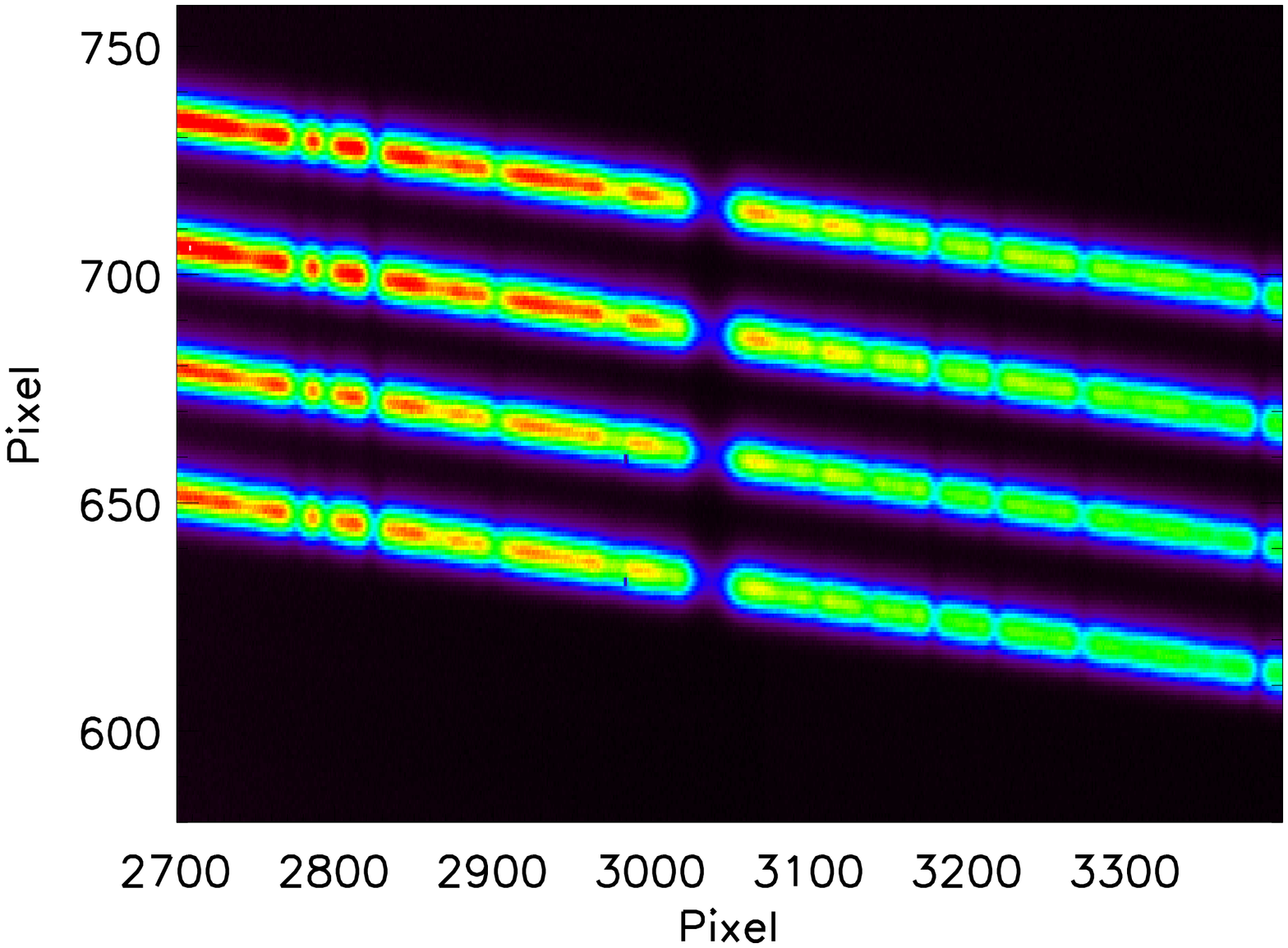}
\includegraphics[width=0.32\linewidth, angle=90]{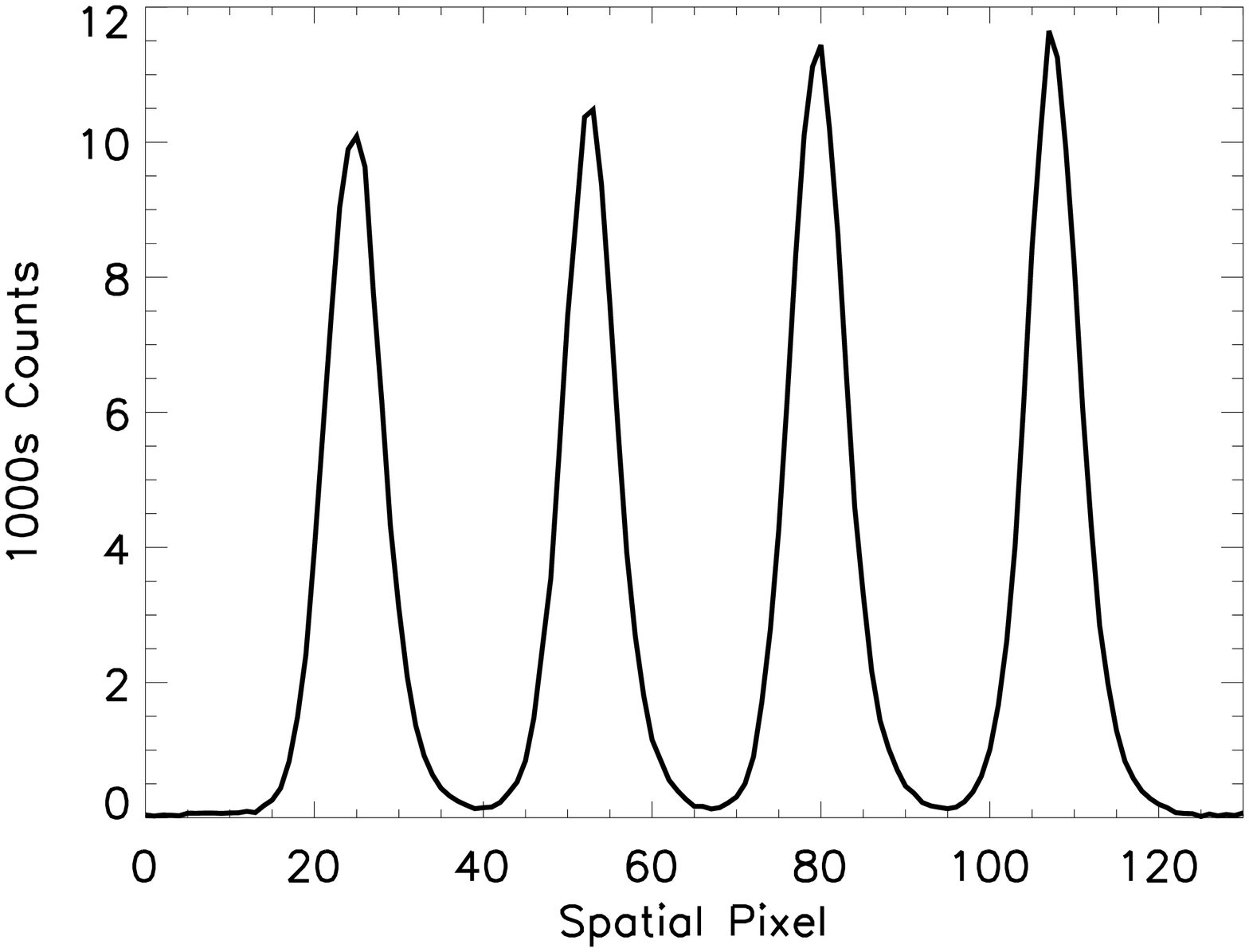}
\includegraphics[width=0.32\linewidth, angle=90]{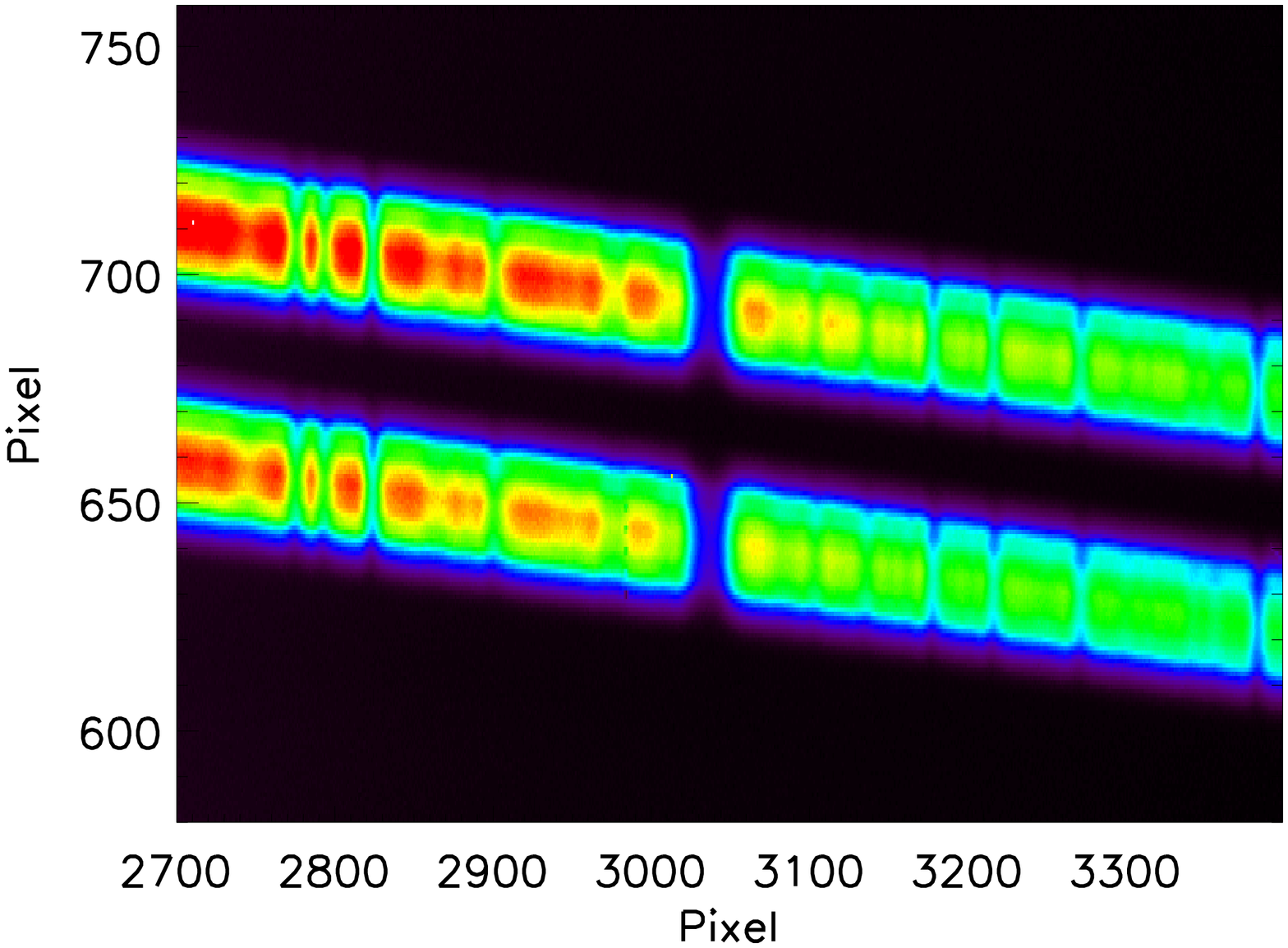}
\includegraphics[width=0.32\linewidth, angle=90]{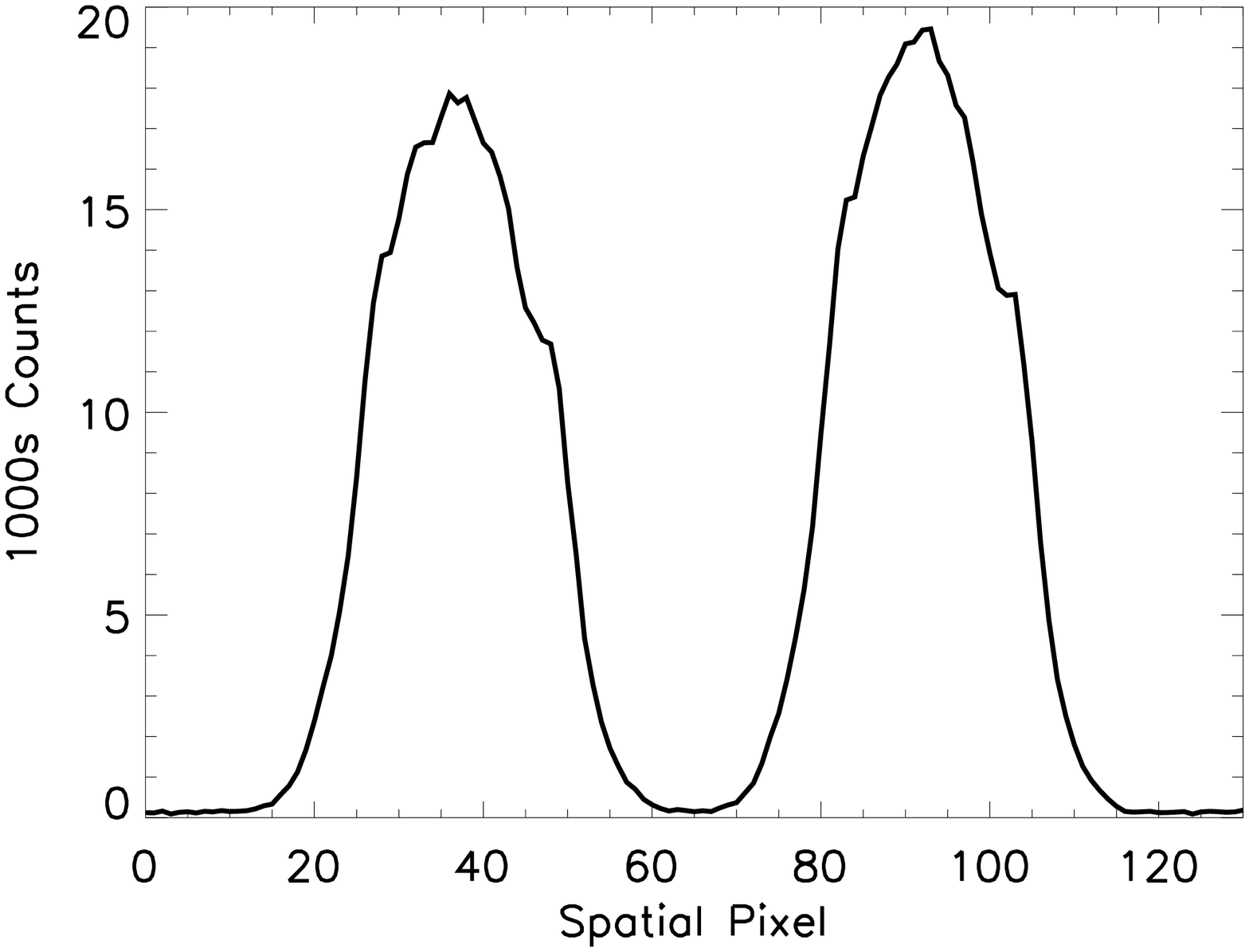}
\caption{ \label{stellar_psfs} This Figure shows examples of data delivered by the different modes available to HiVIS users. Each row shows a different observing mode. The left column shows a small portion of a single exposure with the wavelength direction as generally horizontal and the spatial direction as vertical. The normal dual-beam polarimetric mode is shown in the top row. Two orthogonally polarized orders are imaged on the CCD. The corresponding spatial profiles are computed as a spectral average in the spatial direction across an order. The spatial profile is shown on the top right. Two beams are seen with a FHWM of roughly 10 pixels. The Savart plate analyzer separates the two orthogonally polarized beams by roughly 54 pixels aligned with the spatial direction of the spectral orders. The dekker is set to a slit length less than half this displacement in the spatial direction.  The liquid-crystal charge-shuffled mode is shown in the middle row.  The two distinct liquid crystal modulation settings are inter-woven in 4 beams recorded in a single exposure. We use a charge-shuffling distance of 27 spatial pixels, half of the Savart plate beam displacement.  Two beams are stored for one modulation state in non-illuminated buffer pixels while the photoelectrons are accumulated for the second liquid crystal modulation state. Four beams are recorded using two illuminated spectral regions and adjacent non-illuminated buffer regions.  The smear-mode is shown in the bottom row. The accumulating charge packet from both orthogonally polarized beams is slowly shuffled to near-by non-illuminated rows as the exposure progresses. This effectively increases the full-well capacity of the image and increases the saturation limit of the detector.   }
\end{center}
\end{figure*}

\subsection{HiVIS Polarimeter Upgrades}

	In order to accomplish liquid crystal modulation and synchronous charge shuffling two tasks had to be accomplished:	
\begin{enumerate}
\item Enable clocking of photoelectrons under software control during an exposure.
\item Synchronize charge clocking with liquid crystal modulation state control.
\end{enumerate}

	 This allows us to record the 4 spectra (2 modulation states in a dual-beam optical path) in one exposure. The detector control electronics were upgraded using the STARGRASP hardware and software package. This package was developed for orthogonal transfer applications \citep{Tonry:1997ew, Burke:2007gj, Onaka:2008cc}. This new controller package will move accumulating photoelectrons an arbitrary number of pixels and arbitrary number of times using software commands during an exposure and before detector readout. The charge packet accumulating in each pixel of a CCD column can be clocked both towards and away from the shift register with simple command-line tools. The typical row transfer time depends on several settings but is less than 10$\mu$s for our devices. Since the focal plane of cross-dispersed echelle instruments naturally have separated spatial and spectral directions to the focal plane illumination pattern, this clocking is set to be along the spatial direction of the slit image which is also the readout direction. 

	 Normal CCD's typically do not allow charge transfer in two orthogonal directions but our controller firmware can clock charge away and towards the shift register. By synchronizing these shifts with changes in modulator states, adjacent pixels can be uses as storage buffers without adding detector readout noise. If the order separation is large, the charge can be electronically smeared in the spatial direction to effectively increase the full-well capacity during the exposure. This allows much more efficient use of the pixels in the CCDs. 

	Meadowlark nematic liquid crystal modulators (LCs) were installed and a basic polarization calibration analysis was reported in \citet{Harrington:2010km}. The HiVIS control software was updated to include synchronous command of the modulators and the charge shuffling on the focal plane mosaic. The STARGRASP package uses ethernet communications over a local gigabit network to synchronize external devices with the charge shifting. The timing jitter is less than 10 milliseconds for our ethernet commands. The LCs have a nominal switching time of 40 milliseconds. Thus timing jitter between the charge shuffling and the liquid crystal switching will be small and will typically be averaged over many switching cycles. 

	The last optical requirement was to substantially reduce the slit height with a dekker so the 4 recorded beams were optically separated by our Savart plate analyzer. A dekker width of 30 spatial pixels or less must be used with our Savart plate to ensure optical order separation. The Savart plate displacement is between 53 and 55 pixels thus we charge shuffle by 27 pixels in the default operating modes.

\subsection{HiVIS Spectrograph \& New Observing Modes}

	With the new 4k by 4k mosaic focal plane, we can continuously cover a wavelength range $>$220nm typically in the 550nm to 950nm range with good overlap between spectral orders. The order overlap is usually between 10\% and 40\% depending on the order and cross-disperser settings. The optical design delivers a two-pixel per 0.35`` monochromatic slit image for the highest spectral resolution ($R\sim$50,000) settings. This requires spectral sampling in the range of 4pm to 6pm per spectral pixel. For charge shuffling at high signal-to-noise ratios, we adapted the dekkar to use the 1.5`` slit for this campaign with a spectral resolution of $R=$12,000. We strongly over-sample in the spectral direction with sampling of roughly 4.3pm at H$_\alpha$. This allows us to efficiently observe bright targets to high SNRs without saturation while minimizing several kinds of instrument instability errors. Note ESPaDOnS spectra have quite similar spectral sampling to HiVIS.  ESPaDOnS samples at 2 points per monochromatic slit image for R=68,000 and we sample at 2 points per R=50,000 monochromatic slit image. Since the HiVIS slits are selectable with a filter wheel and our typical science cases do not require high spectral resolution, we used a wide slit for high throughput at R=12,000 while maintaining the same spectral sampling.

	The modulator mechanical package is flexible and can be configured for liquid crystals or rotating achromatic retarders in a few minutes. Additionally, the HiVIS control software has three main observing modes. The first is standard dual-beam polarimetry which only records one modulation state in two beams. The second mode is the liquid crystal modulated charge shuffling technique where four independent spectra are recorded, two for each modulation state. The third is called smear-mode where the accumulating charge packet is slowly shuffled back and forth during an exposure to more evenly distribute the charge in all available spatial pixels. This smear mode is used in bright star campaigns to effectively increase the saturation limit and reduce readout overheads by more fully using the available full well capacity of all available spatial pixels. Figure \ref{stellar_psfs} shows small regions of an image for each observing mode on the left and the corresponding spatial profile on the right. 
	
	There are several advantages gained by using these observing modes for high signal-to-noise observations. Typically bright targets saturate the detector and require exposure times substantially shorter than the readout time. By implementing charge shuffling we double the time spent integrating for a fixed readout time and substantially increase observing efficiency. 
	
	In the case of long-slit instruments, the delivered spatial profile typically takes on the shape of the atmospheric seeing combined with any beam jitter and tracking errors. This profile is normally strongly peaked and Gaussian-like in shape where the core saturates leaving many underexposed spatial pixels. By charge shuffling, this accumulating charge can be more evenly distributed across the focal plane. Beam location changes induced by the guiding and tracking system errors also cause changes in calibration and systematic uncertainties in several parameters. By doing a full dual-beam modulation sequence within a single exposure, many of these effects are mitigated. The liquid crystals are easily tuned via software commands to have different modulation schemes.  
	
	The calibration of the liquid crystals is easily accomplished with our polarization calibration unit shown in Figure \ref{hivis_block} and outlined in \citet{Harrington:2010km}. Additionally, we installed a new dichroic on our slit-viewing guider camera. This allowed faster and higher quality tip/tilt correction using a pupil steering mirror at 1Hz rates for targets brighter than V=8 under nominal conditions.

\begin{figure} [!h, !t, !b]
\begin{center}
\includegraphics[width=1.05\linewidth, angle=0]{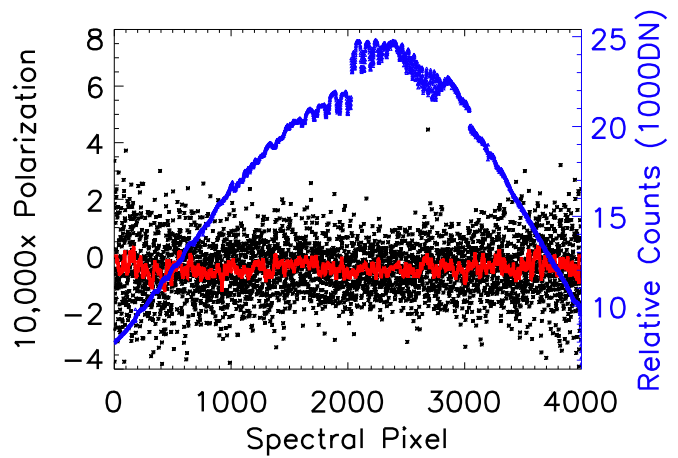}
\caption{ \label{system_stability} This Figure shows the flat field lamp intensity spectrum (blue) and polarization errors (black and red) for a single spectral order. We show the data at our nominal 12,000 spectral resolution and R=100,000 spectral sampling. The right-hand y axis shows the intensity spectrum in detected counts per exposure. The intensity spectrum in blue shows a strong blaze function intensity variation across the order along with the typical CCD mosaic gaps and pedestal offsets. The two CCDs in the mosaic each have two amplifiers with slightly varying gains and biases. The intensity was left uncalibrated to illustrate these CCD effects. Spectral features from the lamp and atmosphere are seen in the middle of the spectrum. The left-hand y axis shows the polarization spectrum multiplied by 10,000.  Amplifier and CCD gaps are not visible in the polarization spectrum. The computed polarization is dominated by shot noise below the 0.01\% level at full spectral sampling. The RMS of the polarization spectrum is substantially decreased by averaging adjacent spectral pixels, as seen in the red curve. This averaging shows that photon statistics are the dominant noise source.  See text for details.}
\end{center}
\end{figure}

\subsection{Liquid Crystal Modulation Rates}

\begin{figure} [!h, !t, !b]
\begin{center}
\includegraphics[width=0.99\linewidth, angle=0]{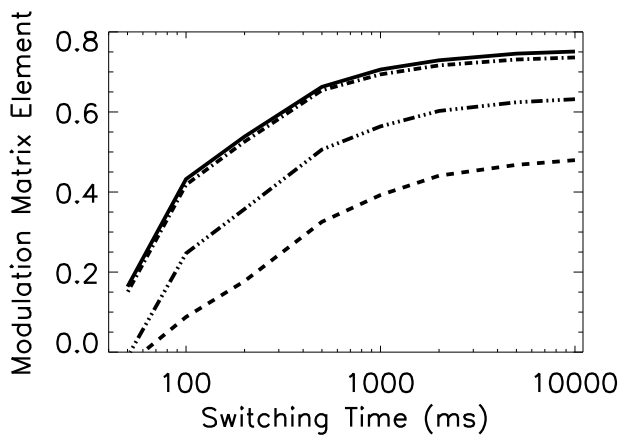}
\caption{ \label{lcvr_efficiency_vs_modrate} The modulation matrix elements as the LCVR and charge-shuffling system is driven at different shuffling speeds.  The Meadowlark LCVRs have a nominal switching time of 20-50ms depending on the temperature, wavelength and retardance change commanded. As the modulation speed is increased the liquid crystals spend a larger fraction of the integration time in intermediate states while switching retardance, thus reducing the efficiency of the polarization modulation. We nominally use a switching time of 1 second or slower to achieve high polarization modulation efficiency with HiVIS. }
\end{center}
\end{figure}

	To determine the optimum liquid crystal modulation rate, the depolarizing effect caused by the nominal 5\% to 95\% liquid crystal retardance switching time of 40 milliseconds speed was tested. Figure \ref{lcvr_efficiency_vs_modrate} shows a few select modulation matrix elements derived as functions of the switching time when using our polarization calibration unit at the HiVIS slit. As expected, Figure \ref{lcvr_efficiency_vs_modrate} shows a general reduction in modulation matrix element amplitudes at switching times faster than 500ms (or a 2Hz modulation frequency). As the liquid crystals are driven faster, they spend a proportionally longer time switching than resting, causing a depolarization of the input signal. For our typical use case, modulation of 2Hz or slower is selected.  Under these conditions the number of cycles is typically low (e.g. 120 shuffles for a 60 second exposure).  

	We have extended calibrations done with LoVIS to the new HiVIS charge-shuffled liquid crystal modulated configuration. As a standard non-precision tuning, we set our liquid crystal voltages to those derived in the laboratory for nominal H$_\alpha$ modulation using a Stokes definition scheme where each exposure measures a single Stokes parameter $QUV$. With the polarization calibration unit, we inject six pure $\pm$ $quv$ Stokes parameters we can derive two redundant sets of demodulation matrices. With this information, we also can extract the chromatic effects of the calibration quarter wave plate as well as the stability of the system. 

	The derived liquid crystal modulation matrices are quite stable over the course of a night. Each night we observe, we collect two complete modulation matrices derived using the $+$ and $-$ input states. Experiments done many times show that independent sets of calibration observations can be used to demodulate other sets of observations to within the shot noise of $<$0.001 in any individual Stokes parameter. When a demodulation is done using the same input state, say $+$ Stokes inputs demodulating measurements of $+$ Stokes inputs, the residuals are always shot-noise limited.  When demodulation is done using opposite input states, the dominant effect is the chromatism in the calibration quarter wave retarder.  This effect is removed when using the daytime sky calibrations to calibrate the cross-talk in the demodulated spectra.

\subsection{Polarimeter Summary}

	With these optical improvements, we performed several tests to demonstrate the performance of this new hardware. In summary, the liquid crystals achieve $>$95\% modulation efficiency at speeds of 1Hz or slower and the calibrations are stable over several-day timescales. The scattered light, bias stability and calibration data all give consistent results. In a series of 400 exposures, we were able to demonstrate the repeatability of the polarization calibrations down to the 0.01\% SNR limit at full spectral resolution across the spectrum in the presence of molecular band features. Figure \ref{system_stability} shows polarization calibrations where a set of 200 images was used to calibrate another 200 independent Stokes $q$ exposures. The results are stable with a signal-to-noise ratio (SNR) greater than 10,000 at full spectral sampling and resolution. Figure \ref{system_stability} shows the intensity spectrum across a single spectral order along with the corresponding residual error between the two independent 200-image calibration sets.  There is a strong blaze function and some spectral absorption features from the lamp and atmosphere seen. Even with these intensity variations, spectral smoothing does not reveal underlying systematic effects that limits the instrument.  All spectral orders show only statistical noise when spectrally binned by factors up to 128.  The corresponding calibrations obey a $\sqrt{N}$ behavior down to 0.001\% levels. As an example in Figure \ref{system_stability}, the red line shows a simple boxcar smoothing and a corresponding reduction in statistical noise of the spectra. Liquid crystal instabilities, calibration inaccuracies or other detector-based effects do not appear. We conclude that our coud\'{e} instrument does not have substantial systematic errors that limit the calibration accuracy. Our instrumental artifacts are below 0.01\% levels at full spectral sampling across the full spectral range for internal calibration sources.

\section{Calibrations: Stellar Standards}

	 Observations of unpolarized and polarized standard stars are used to derive instrument calibration stability and precision limits. With the ability to record 4 modulated beams in a single exposure with 1Hz modulation rates, we can reduce the continuum polarization errors caused by instrument instabilities and atmospheric changes on timescales of seconds to minutes.
	 
\begin{figure} [!h, !t, !b]
\begin{center}
\includegraphics[width=0.75\linewidth, angle=90]{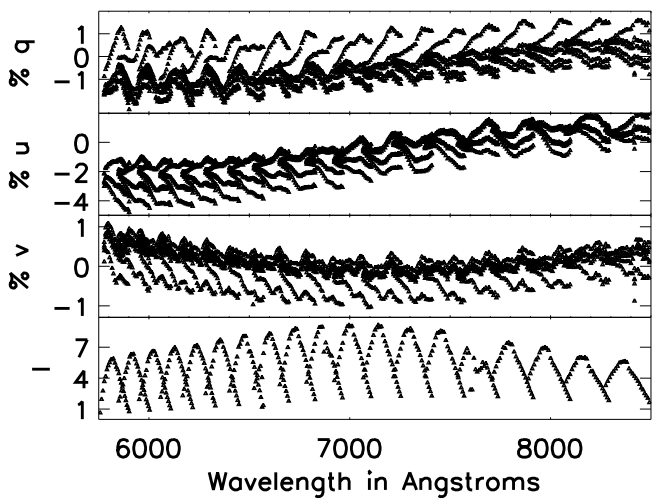}
\caption{ \label{inducedpol_unguided} This Figure shows the computed polarization spectra for repeated measurements of the unpolarized standard star HR2943. Slit-guiding and charge-shuffling was disabled to demonstrate the typical effects introduced by atmospheric transmission variation (clouds), telescope jitter, guiding errors and other instrumental errors on the continuum polarization. The intensity spectrum in the bottom row shows all 19 extracted spectral orders and the associated blaze functions. Each $quv$ continuum polarization measurement shows substantial variation of over 2\%. There are two components to the $quv$ spectra: the spectrograph calibration function and the telescope induced polarization. See text for details.}
\end{center}
\end{figure}

	 HiVIS observations of unpolarized standard stars without guiding or charge shuffling have shown the derived continuum polarization to have high variance of up to 5\%. Several unpolarized standard star lists are available in these references \citet{1974psns.coll.....G, Fossati:2007ud, GilHutton:2003di}.  There is dependence on telescope focus and sensitivity to strong wind-induced telescope pointing jitter. We find the origin of this continuum polarization variance comes from changes in throughput between both polarized beams produced by the Savart plate analyzer. Instabilities in the telescope pointing, guiding and tracking when combined with intra-order contamination between independent modulated beams induce time-dependent continuum changes. In this configuration there is a mostly static spectrograph calibration function in addition to the induced polarization. As an example, Figure \ref{inducedpol_unguided} shows repeated observations of the unpolarized standard star HR2943 recorded without using tip/tilt slit-guiding or charge-shuffling \citep{1974psns.coll.....G}. We performed the typical {\it Stokes definition} modulation scheme for beam-swapping used by most night-time spectropolarimeters. This consists of 6 exposures with the modulators tuned such that two modulation states are dedicated to measuring a single Stokes parameter. Figure \ref{inducedpol_unguided} shows $quv$ in the top three panels. The corresponding detected count levels are shown in the bottom panel. There are substantial spurious variations in $quv$ without any systematic temporal trends.

\begin{figure} [!h, !t, !b]
\begin{center}
\includegraphics[width=0.75\linewidth, angle=90]{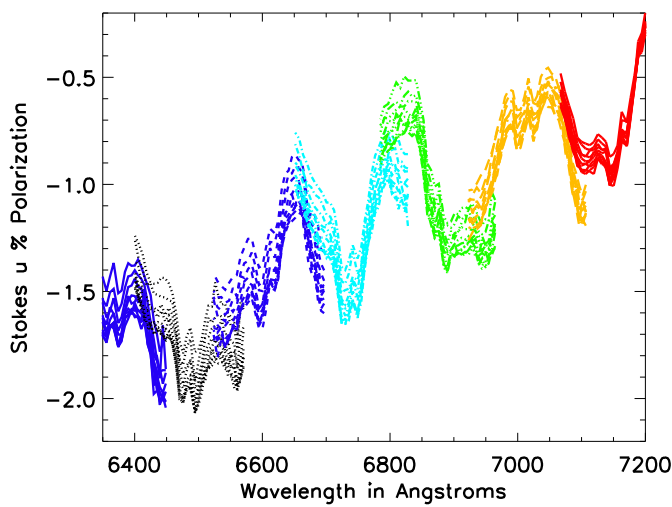}
\caption{ \label{unpolstd_continuum_U_jan10} The computed Stokes $u$ continuum polarization for the unpolarized standard star HD 38393 observed January 10$^{th}$ 2013.  Spectral averaging by a factor of 80x for high SNR has been applied. At this spectral binning there are 50 spectral points per spectral order. Each spectral order is shown in a different color for this wavelength range. The different colors nearly overlap but with 0.1\% offsets. The intra-order and inter-order variation is a stable function representing the $\sim$0.5\% variation across each spectral order. The telescope induced polarization causes the curves to change slowly with time and with wavelength. The total time elapsed during this data set was 92 minutes (09:59z to 11:31z). The statistical noise limits are smaller than 0.05\% per pixel at this spectral binning.}
\end{center}
\end{figure}

\begin{figure} [!h, !t, !b]
\begin{center}
\includegraphics[width=0.75\linewidth, angle=90]{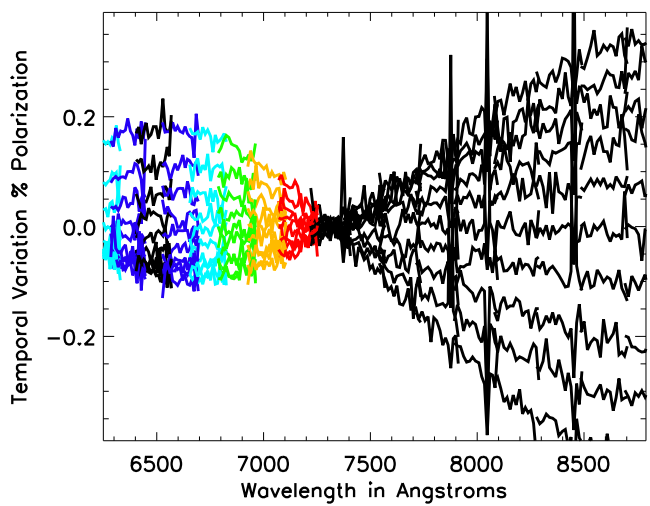}
\caption{ \label{unpolstd_continuum_Ures_jan10} This Figure shows the residual Stokes $u$  temporal variation for the 10 sequential HD 38393 observations from Figure \ref{unpolstd_continuum_U_jan10} observed January 10$^{th}$ 2013. Each spectral order is a different color. The variation with time is a smooth function and all spectral orders overlap for all exposures.  See text for details.}
\end{center}
\end{figure}

	When HiVIS charge-shuffling is enabled, a complete polarimetric data set consists of 3 exposures and 6 modulation states delivering 12 intensity modulated spectra. When recording two modulations states in a single exposure with the new guiding system enabled, stable continuum and spectrograph calibration measurements are obtained. Figure \ref{unpolstd_continuum_Ures_jan10} shows a sequence of 10 Stokes $u$ polarization measurements obtained over 1.5 hours. We observed the unpolarized standard star HD 38393 with 4Hz modulation and nominal H$_\alpha$ Stokes definition liquid crystal voltages  \citep{1974psns.coll.....G}. The standard data reduction scripts were used to compute the polarization. The data in Figure \ref{unpolstd_continuum_U_jan10} have been spectrally binned by a factor of 50 to reduce the statistical (shot) noise limits below 0.05\%. A consistent $\sim$0.5\% variation remains across individual spectral orders that comes most likely from beam throughput variations and intra-order beam contamination. This variation we call the spectrograph calibration function. Each color shows a different spectral order. There are discontinuities at orders boundaries from changes in optical transmission across a spectral order. However, these discontinuities are now stable and can be calibrated. The computed continuum polarization includes any differential instrumental throughput variations, since the continuum polarization is calculated using a simple normalized double-difference:

\begin{equation}
q= \frac{Q}{I} = \frac{1}{2}(\frac{a-b}{a+b} - \frac{c-d}{c+d})
\end{equation}
	
	This normalized double difference method is quite common and can easily be compared with other difference and ratio methods \citep{Harrington:2008jq, Bagnulo:2009bz, Bagnulo:2012dr, Patat:2006en, 1992SPIE.1746...22E}.  With 4Hz modulation and a single-exposure per polarization spectrum, stable variations with time are seen. Figure \ref{unpolstd_continuum_Ures_jan10} shows the residual temporal changes for the 10 Stokes $u$ measurements of HD 38393. The average $u$ spectrum has been subtracted from each individual $u$ measurement. Each color shows a different spectral order and there is continuity across order edges. The intra-order variation was effectively constant and has been removed. Smooth wavelength dependent changes of $<$0.3\% per hour over the 1.5 hour period are expected as the telescope tracks the star. This contrasts with the erratic 5\% variation seen in Figure \ref{inducedpol_unguided} without charge shuffling or guiding.

	We measured the continuum polarization stabilization as a function of modulation speed. The dichroic-assisted tip-tilt guiding at 0.1Hz was enabled. A single exposure was used to measure the continuum polarization spectrum. In order to isolate the optical stability of the system, the liquid crystal modulators were set to a constant voltage so that all artifacts are caused entirely by optical instabilities with minimal impact from real polarization changes induced by tracking the star. The target was observed for 8 to 12 exposures at a single modulation frequency and the root-mean-square (RMS) continuum polarization variation was computed. This RMS was measured at modulation frequencies of 1/60, 1/15, 1, and 4 Hz with a constant 120 second exposure time. The spurious instrumental continuum polarization variation is 1-4\% when doing normal exposure-time-limited observations.  With modulations at 1-minute scales (1//60Hz) and the 4 charge-shuffled beams recorded in a single exposure, the continuum stabilizes to the 0.1\% level.  When modulation rates are increased to the 1Hz range, continuum polarizations were stable to the 0.03\% level. 

	The optimal modulation speed is a trade off between minimizing spurious effects of optical instabilities and other sources of instrumental error.  As the modulation speed increases, the system spends a larger fraction of the integration time in the process of switching states. This reduces the amplitude of the detected polarization signal. A key conclusion is that modulations speeds faster than 1Hz show degraded polarization amplitudes. Additionally, for longer exposures there are more shuffles per exposure. Errors caused by the detector including hot pixels, charge traps and charge transfer inefficiency increase.

\subsection{Polarized Standard Stars}

	The distribution of polarized standard star reference targets is quite uneven and highly clustered at specific locations on the sky. It is often not the case that a polarized standard and an unpolarized standard are available at the same telescope pointing as your science targets in the same season. Spectropolarimetric standards are nearly non-existent and interpolation from Serkowski-law fit parameters is subject to errors. 

	We observed a polarized standard star, HD 30657 with a known linear continuum polarization of 3.72\% in an R filter to demonstrate the calibration process \citep{Whittet:1992gr}. The unpolarized standard star with the closest telescope pointing during our allocated observing time was HD 38393.  Unfortunately, the HD 38393 unpolarized standard star observations were not recorded at an identical telescope pointing to our polarized standard HD 30657.  The unpolarized standard star spectra do not have any intrinsic polarization and hence record only the telescope induce polarization and spectrograph calibrations.  The unpolarized standard star observations measure the $I$ to $QUV$ terms of the Mueller matrix. These terms are nominally stable in time and can be subtracted.  However, the telescope pointing was not identical for the unpolarized and polarized standard star measurements.  Thus, the subtracted induced polarization will suffer from a small unknown offset. 
	
\begin{figure} [!h, !t, !b]
\begin{center}
\includegraphics[width=0.75\linewidth, angle=90]{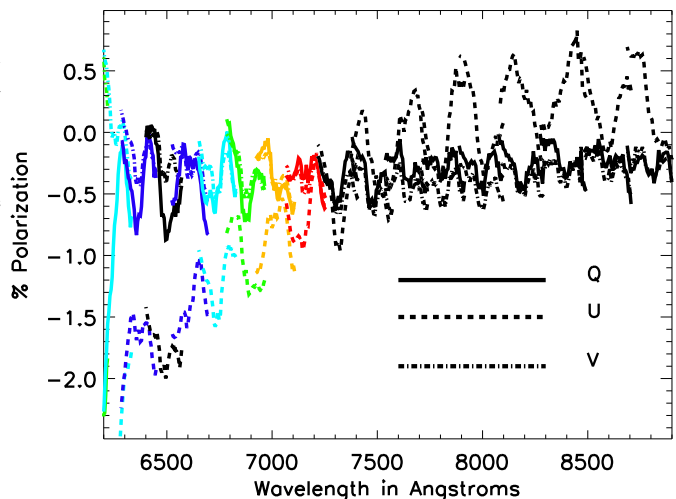}
\caption{ \label{unpolstd_continuum_jan10} The average $quv$ continuum polarization calculation for the unpolarized standard star HD 38393 observed January 10$^{th}$ 2013. The $quv$ spectra are shown over the entire HiVIS wavelength range. The dominant factor is the typical spectrograph calibration in the charge shuffling configuration. There were 10 complete measurements recorded at 4Hz modulation rates recorded and then averaged. Certain spectral orders are shown in color for direct comparison with Figures \ref{unpolstd_continuum_Ures_jan10} and \ref{unpolstd_continuum_U_jan10}.  See text for details.}
\end{center}
\end{figure}

	The typical change in telescope induced polarization with pointing changes was shown in Figure \ref{unpolstd_continuum_U_jan10} and is fairly small. As shown in Figure \ref{unpolstd_continuum_Ures_jan10} the $u$ continuum polarization changes by roughly 0.2\% over roughly one hour of tracking. The average $quv$ spectra for the HD 38393 unpolarized standard star are shown in Figure \ref{unpolstd_continuum_jan10}. The spectrum is dominated by $u$ polarization and spectrograph intra-order variations.

	We simply subtract the HD 38393 unpolarized standard star calibration measurements shown in Figure \ref{unpolstd_continuum_jan10} from observations of the polarized standard star HD 30675 taken on the same night at a nearby telescope pointing. This allows us to remove the $I$ to $QUV$ artifacts within the uncertainty of the different telescope pointings. The resulting $quv$ spectra from HD 30675 are thus smooth functions of wavelength calibrated for $I$ to $QUV$ effects.
	
\begin{figure} [!h, !t, !b]
\begin{center}
\includegraphics[width=0.75\linewidth, angle=90]{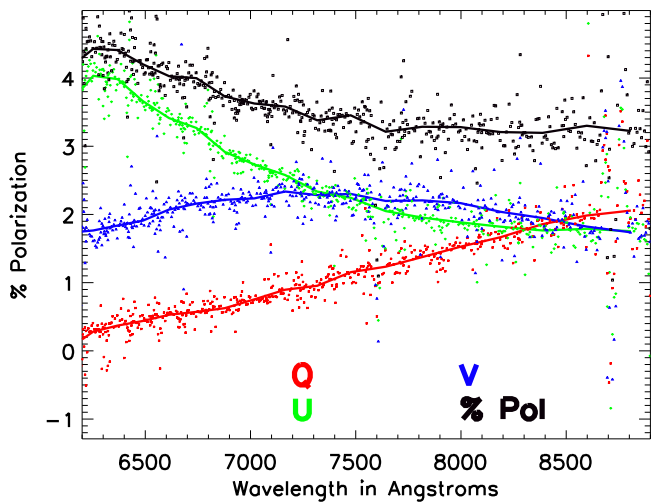}
\caption{ \label{polstd_continuum_jan10} The continuum polarization for a polarized standard star HD 30657 observed on January 10$^{th}$ 2013. In Whittet et al. 1992 this star is quoted to have an R-band linear polarization of 3.72\%. Stokes $q$ is shown in red, $u$ in green, $v$ in blue and the total degree of polarization in black. The colored symbols show the spectrum rebinned to 20 points per spectral order to highlight noise levels and any order edge mismatches. The solid colored lines show the spectra plotted with one point per spectral order.  The telescope induced polarization and associated continuum errors have been removed from this measurement using an unpolarized standard star observed at a nearby pointing.  Observations were recorded from 11:48 to 12:46.  The telescope pointing varied from elevation,azimuth of (35,291) to (22,294).  Two sets were recorded with 300 shuffles per exposure and a 1Hz modulation frequency. This corresponds to 300 seconds per beam, four beams per exposure with a total of three exposures. }
\end{center}
\end{figure}

	Figure \ref{polstd_continuum_jan10} shows these calibrated measurements as the colored symbols with red as $q$, green as $u$ and blue as $v$.  The degree of polarization ($p$) is also shown as the black symbols. The points have been spectrally binned by a factor of 50 to decrease the statistical noise on this V=7.6 star. There is strong cross-talk from this linearly polarized standard star as the $v$ spectrum is a nearly constant 2\% with substantial changes with wavelength.

	The degree of polarization (black symbols) measured for HD 30675 in Figure \ref{polstd_continuum_jan10} closely matches the 3.72\% R-band polarization of \citet{Whittet:1992gr}. As the polarized and unpolarized star observations were not taken at the same telescope pointing, there is uncertainty in the telescope induced polarization. There is little variation across spectral orders seen in Figure \ref{polstd_continuum_jan10}. The predominant effects seen in continuum polarization is shot noise and the telescope induced cross-talk. With these newly-stabilized continuum polarization measurements, we can now assess our telescope calibrations with much higher accuracy.

\subsection{Standard Star Summary}

	We have measured polarized and unpolarized standard stars at resolutions of 12,000 and spectral sampling of 100,000 to demonstrate calibration of the spectropolarimeter and proof of the data analysis algorithms. The calibrations were spectrally averaged to 50 points per spectral order for an effective 1-point sampling of R$\sim$1250. The calibrations are smooth functions of wavelength and show well resolved smooth variation across individual spectral orders. We have demonstrated stable continuum polarization calibrations at these spectral samplings. Modulation at speeds of 1Hz give stable measurements of both polarized and unpolarized standard stars below 0.1\% precision levels when spectrally binned to high SNRs. This system substantially reduces the types of instrumental effects when comparing multiple lines across the spectrum. We find that relatively slow modulation rates combined with accurate guiding are sufficient to stabilize the optical path of the instrument and remove effects from pointing jitter. 

\section{Calibrations: Daytime Sky Validation}

	Measuring absolute polarization with HiVIS requires a full-Stokes polarization calibration process to account for all the telescope and instrument induced cross-talk. In coud\'{e} path instruments and complex systems, this absolute calibration is often the most unstable and difficult. The daytime sky is a very bright, highly polarized and easily modeled calibration source. It provides linear polarization at any telescope pointing, has similar surface brightness to bright stellar targets and does not waste night time observing. We have developed techniques to observe the daytime sky polarization on our low resolution spectrograph, LoVIS in order to derive polarization calibrations for the telescope and instrument \citep{Harrington:2011fz}.  In \citet{Harrington:2010km} we presented initial calibrations of the HiVIS liquid crystal polarimeter. Here we have successfully applied this daytime sky techniques to HiVIS and the liquid crystal demodulation algorithms to our new charge shuffling instrument. Simple procedures can project these measurements to all telescope pointings and calibrate science observations.

\begin{figure} [!h, !t, !b]
\begin{center}
\includegraphics[width=0.75\linewidth, angle=90]{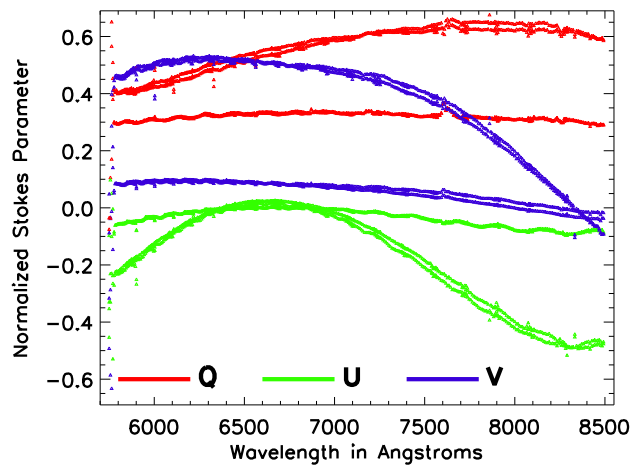}
\caption{ \label{demodulated_daytime_sky_spectra} The demodulated daytime sky polarization spectra on May 5th 2012.  The telescope pointing was azimuth 360 and elevation 90. Data was taken at two different times (2.62 and 4.86 hours UT). Each measurement has two independent demodulation calibrations applied with the two calibration unit demodulation matrices. The curves largely overlap showing minimal chromatic effects from the polarization calibration unit. Spectral averaging by a factor of 32 was applied to 125 data points per spectral order to achieve high SNR.  The consistency of the data inside each spectral order as well as agreement in regions of spectral order overlap is seen in the smooth spectral variation of the $quv$ spectra.}
\end{center}
\end{figure}

	In order to derive the polarization calibrations and the system Muller matrix, full-Stokes observations must be obtained of the daytime sky polarization spectra over a span of typically a few hours. To obtain demodulation matrices, observations of calibration lamps taken with our polarization calibration injection unit are also recorded and processed. This unit consists of a rotating wire grid polarizer and a rotating achromatic quarter wave plate. The measured daytime sky polarization spectra are demodulated with our standard procedures using the polarization calibration unit optics. Note that for error checking, we record two completely independent demodulation matrices using our polarization calibration stage. All polarization calibration unit optics are rotated by 180$^\circ$ so that we have two independently calibrated demodulated spectra for comparison.

	Examples of demodulated daytime sky polarization spectra are shown in Figure \ref{demodulated_daytime_sky_spectra}. The spectra have been binned to 125 measurements per spectral order (from an original 4000) in order to highlight systematic errors. The statistical noise is smaller than the width of the line used for plotting. One can see in the Figure that all curves overlap reasonably well and show only small intra-order effects. There are two curves per Stokes parameter corresponding to the two redundant demodulation matrices. To demonstrate the linear to circular cross-talk induced by the telescope, observations at two separate times are shown. The sun moved substantially during the 2.2 hours elapsed between observations so that the input polarization angle to the telescope is known with two values. There is a substantial amount of circular polarization (blue) measured at short wavelengths caused by telescope cross-talk. One can see a substantial variation in the measured $q$ and $u$ amplitudes with wavelength indicating the telescope induced rotation of the linear polarization orientation.

	In order to derive the telescope calibrations as a rotation in $quv$ space, the procedures outlined in \citet{Harrington:2011fz} are followed. The degree of polarization is computed as $\sqrt{q^2 + u^2 +v^2}$ for each spectrum after demodulation. Each measurement is then scaled to 100\% polarization which places the measurement on the Poincar\'{e} sphere. The degree of polarization expected for simple Rayleigh scattering in the atmosphere follows a simple relation whose range of validity has been experimentally verified at many sites \citep{Coulson:1980wq, Cronin:2006jy,Cronin:2005kd, Lee:1998hj, Liu:1997dg, North:1997hk, Pomozi:2001tn, Pust:2006fd, Pust:2009fq, Suhai:2004ed, Voss:1997ix, Swindle:2014wc}.
	
\begin{figure} [!h, !t, !b]
\begin{center}
\includegraphics[width=0.75\linewidth, angle=90]{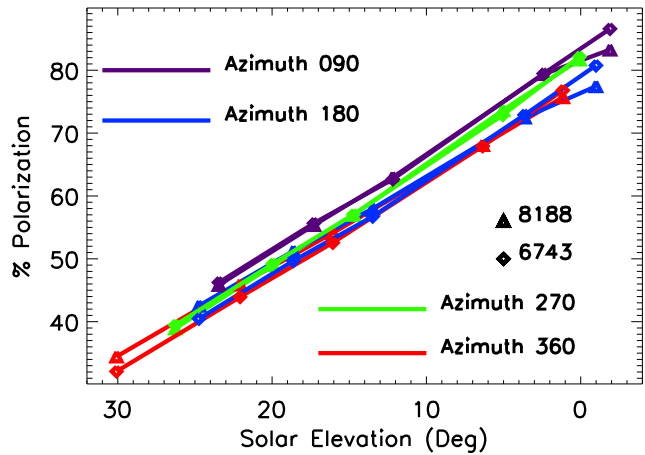}
\caption{ \label{daysky_poldegree} The measured degree of polarization for the HiVIS daytime sky polarization spectra. The telescope was pointed at the zenith (elevation=90$^\circ$) as the telescope azimuth was changed between [90, 180, 270 and 360] as a sequence in time. The different symbols show two different spectral orders: 6700{\AA} as diamonds and 8200{\AA} as triangles. Different colors represent different telescope azimuths as outlined in the legend. As the sun sets, the Zenith degree of polarization is expected to rise nearly linearly as the solar elevation angle decreases.}
\end{center}
\end{figure}
		
	The degree of polarization for a typical set of measurements when pointed at the Zenith is shown in Figure \ref{daysky_poldegree}. The linearly increasing trend in degree of polarization is expected as the scattering angle (sun-zenith-telescope) increases. The slight difference in polarization between East (090) and North (360) azimuths reflects telescope depolarization as well as errors in demodulation. The measurements in Figure \ref{daysky_poldegree} are consistent with that Rayleigh sky model and a very high (85\%) maximum degree of polarization. These values are typical and are similar to all-sky polarization measurements of Haleakala, Mauna-Loa observatory, and other sites \citep{Dahlberg:2011wk, Dahlberg:2009jh, Swindle:2014wc}. The high measured degree of polarization shows that HiVIS with the liquid crystals and charge-shuffling on AEOS has little depolarization, certainly less than 5\%. As with LoVIS and our other observing modes, observations of a wire-grid polarizer covering the calibration sources at the coud\'{e} room entrance show polarization detected above 98\% and high instrument detection efficiency.

\subsection{Stability and Accuracy of Calibrations}
	
	Our daytime sky calibration technique only requires a minimum of two observations to derive a Mueller matrix estimate \citep{Harrington:2011fz}. Additional observations provide redundant information that can be used to test systematic errors and improve accuracy. Independent calibrations can be derived using data sub-sets to assess calibration systematic errors. A large dataset of daytime sky observations was recorded with HiVIS using charge-shuffling between April and June 2012 for verification. Calibrations were recorded at cardinal pointings (North, East, South and West azimuths) at elevations of 90$^\circ$, 70$^\circ$ and 55$^\circ$.  We collected 4 to 9 measurements at many telescope pointings and repeated observations over multiple epochs. For our statistical analysis, we chose data sets with the telescope pointed at the zenith for the cardinal azimuths recorded on April 21$^{st}$, 22$^{nd}$ and May 10$^{th}$ 2012. For each observation date, an independent system Mueller matrix was computed following \citep{Harrington:2011fz}.

\begin{figure} [!h, !t, !b]
\begin{center}
\includegraphics[width=0.75\linewidth, angle=90]{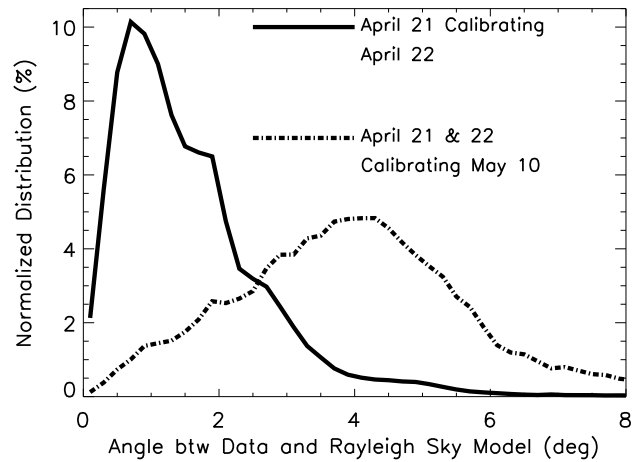}
\caption{ \label{derotated_data_histogram} A histogram of the residual rotation angle difference between theory and calibrated HiVIS daytime sky polarization spectra. The angular difference between theoretical Rayleigh sky input polarization and the calibrated HiVIS measurements on the Poincar\'{e} sphere is computed.  There are 19 separate spectral orders spectrally averaged to 50 points per spectral order used to create the Histograms. Calibrations were derived using April 21st 2012 data and applied to observations from April 22nd 2012 giving residual angular error of 1$^\circ$ to 2$^\circ$ between theoretical sky models and calibrated observations.  Calibrations were also derived using both April 21 and 22nd 2012 data which were applied to May 10th 2012 observations.  In this case, the residual angular error is more typically in the range of 3$^\circ$ to 6$^\circ$.}
\end{center}
\end{figure}

	 Since this calibration method effectively applies rotations on the Poincar\'{e} sphere, we found it instructive to show the rotational error between the calibrated observations and the theoretical daytime sky polarization. First we compute the demodulated, fully polarized observations. We then apply our three separate system Mueller-matrices to calibrate all the sky measurements. This gives us three sets of data calibrated using three independent system calibrations. The residual angular distance on the Poincar\'{e} sphere between these calibrated observations and the theoretical Rayleigh sky polarization is computed for each wavelength and telescope pointing.

	Figure \ref{derotated_data_histogram} shows the resulting histograms of these residual angles. Each histogram curve has been normalized by the total number of data points so the relative fraction can be easily seen. The solid curve shows a histogram computed using only April 21$^{st}$ and 22$^{nd}$ data sets and calibrations. When calibrations and observations are taken within a 24 hour period, the residual angles are quite small, typically around 1$^\circ$ rotational error in $quv$ space. For this distribution, 50\% of the points are within 1.3 $^\circ$ and 90\% are within 3.1$^\circ$. When the May 10$^{th}$ data is used to calibrate April observations and vice versa, the residual error increases. The dashed line in Figure \ref{derotated_data_histogram} shows calibrations applied after twenty days of elapsed time and a reinstallation of the polarimeter. This histogram shows a peak around 4$^\circ$ rotational error. This distribution has 50\% of the points within 3.9$^\circ$ rotational error however 90\% of the points are still within 6.5$^\circ$.

	There are several potential causes for calibration error. First, the Rayleigh sky theory can show angular variations between actual observations at high precision. Most all-sky imaging polarimetry studies have focused on the degree of polarization but the angle of polarization is not monitored as precisely \citep{Voss:1997ix,Vermeulen:2000kq,Suhai:2004ed,Shaw:2010jw,Pust:2005hl,Pust:2006gc,Pust:2007fl,Pust:2008bj,Pust:2009fq,Pust:2006fd,North:1997hk,Liu:1997dg,Lee:1998hj,Dahlberg:2009jh,Dahlberg:2011wk,Cronin:2005kd,Cronin:2006jy,Coulson:1980wq}.  Recent experiments with a new all-sky imaging polarimeter on Haleakala adjacent to AEOS have investigated this uncertainty \citep{Swindle:2014wc,Swindle:2014ue}.  They find that the angle of polarization can vary by up to 3 degrees in regions of the daytime sky with degree of polarization lower than about 15\%.  Provided we only use daytime sky measurements in regions with measured degree of polarizations above 15\%, this angular variation is smaller than 1$^\circ$ in input $qu$ orientation. Thus, we only use daytime sky measurements with high degree of polarization and minimize this source of uncertainty. 

	Thermal stability of our liquid crystal retarders is another error source. Though we are in a coud\'{e} room, small changes in temperature during the time between modulation matrix measurement and sky observation can cause errors. Self-heating from the control signal is also known to cause retardation changes though this effect is small in our devices. An example can be seen in Figure 5 of \citet{Harrington:2010km}. A small temperature change between April and May could easily cause an instability. Replacement of our liquid crystals with the active temperature stabilization version and better thermal control of the instrument could possibly improve our calibration stability.

\begin{figure} [!h, !t, !b]
\begin{center}
\hbox{
\hspace{-2.0em}
\includegraphics[width=0.95\linewidth, angle=90]{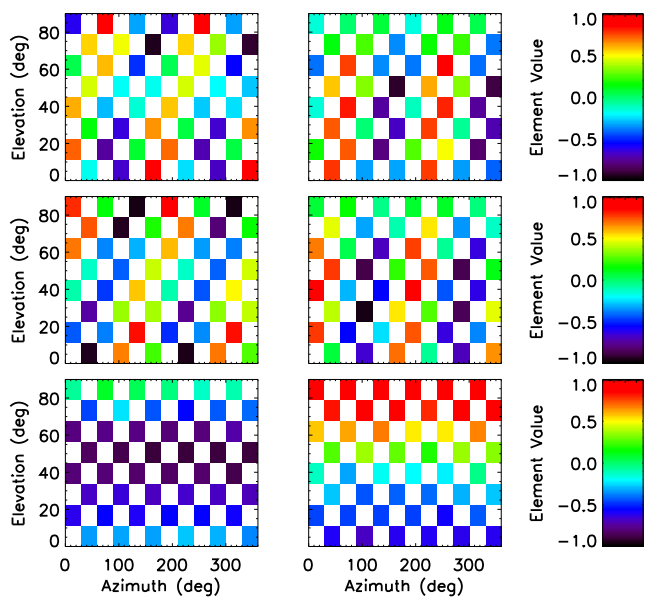}
}
\caption{ \label{mueller_estimate_map} This Figure shows the six Mueller matrix estimates derived via our least-squares method using all data from our 2013 to 2014 calibration efforts. Each colored point represents an altitude-azimuth telescope pointing where we have quality data for estimating the telescope Mueller matrix. The $QQ$ and $QU$ terms are on top. The $UQ$ and $UU$ terms are in the middle. The $QV$ and $UV$ terms are on the bottom. Note the bottom row represents linear to circular polarization cross-talk and the amplitudes are at 100\% for many pointings.}
\end{center}
\end{figure}
\vspace{-1cm}

\subsection{Altitude-Azimuth Dependence}

	Telescope polarization calibration must be applied to observations at all telescope pointings. Creating a telescope Mueller matrix at every altitude and azimuth requires sufficient calibration data to accurately characterize the behavior of the telescope Mueller matrix. An observing campaign was performed during winter of 2013 and summer of 2014 to make a densely covered set of calibrations with HiVIS and charge shuffling. Daytime sky polarization observations were obtained in October, November and December of 2013 as well as in May of 2014.  We recorded many redundant exposures at azimuths of [030, 060, 090, 120, 150, 180, 210, 270, 330, 360]. The corresponding elevations were: [10, 20, 25, 35, 50, 60, 75, 89]. Recent upgrades to the AEOS telescope now allow observations all day. Previously, our telescope pointing elevation limit was above 55$^\circ$ and the sun had to be below elevations of 45$^\circ$.

	The procedure outlined in \cite{Harrington:2011fz} allows one to estimate Mueller matrix elements from Rayleigh sky observations. We scale all HiVIS daytime sky observations to 100\% degree of polarization before we calculate the cross-talk terms of the Mueller matrix. Measurements are put on to the Poincar\'{e} sphere before deriving the best-fit rotation matrices as this results in constructing a physically realistic Mueller matrix. This procedure models telescope as a simple rotation incoming $quv$ polarization to some other $quv$ basis set. As an example of the altitude-azimuth behavior, we show the Mueller matrix estimates for spectral order 3 in Figure \ref{mueller_estimate_map}. 

	 For AEOS, the functional dependence with altitude and azimuth follows simple trigonometric functions and can be fit with a few independent parameters. The dominant polarization cross-talk is caused by rotation of the fold mirrors along the altitude and azimuthal axes. With simple functional coefficients, we can re-compute the azimuthal dependence at an arbitrary telescope pointing. Figure \ref{mueller_matrix_altazi} shows the result of this interpolation. The cross-talk elements of the telescope Mueller matrix are shown for all altitudes and azimuths. This process allows individual Mueller matrix elements to be obtained at the pointing of the astronomical target. After interpolation, the user can derive the best-fit rotation matrix using the standard Euler-angle procedure outlined in \cite{Harrington:2011fz}. This ensures physicality of the telescope Mueller matrix and reduces errors in Mueller matrix term estimation. By representing the telescope Mueller matrix as a rotation matrix, we ensure physicality of the transformation and we do not amplify certain kinds of systematic noise \citep{Kostinski:1993uh, Takakura:2009bq, Givens:1993cl}. 

\begin{figure} [!h, !t, !b]
\begin{center}
\includegraphics[width=1.05\linewidth, angle=0]{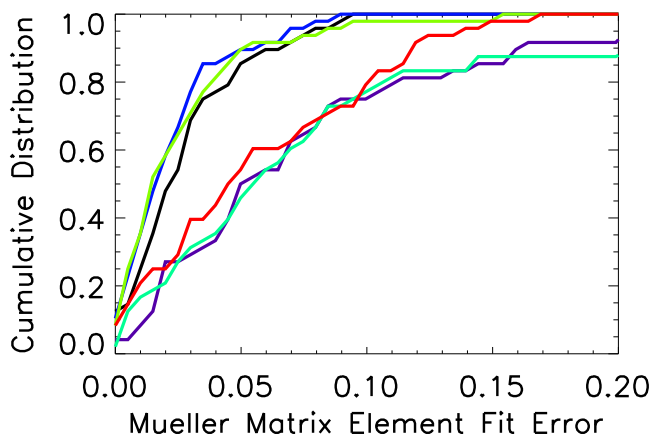}
\caption{ \label{sincos_cumulative_error_distribution} The cumulative error distribution functions when fitting rotation matrix functions of azimuth and elevation to the individual mueller matrix elements. The distribution is based on the difference between the best-fit rotation matrix and each individual least-squares based Mueller matrix estimate at every measured azimuth and elevation. Each color corresponds to one of the six Mueller matrix estimates computed using our daytime sky method. The 68\% confidence interval falls between errors of 0.04 for some elements and 0.10 for other elements.}
\end{center}
\end{figure}

	To quantify the errors, we derive the cumulative distribution function for the difference between the six independent Mueller matrix estimates and the corresponding best-fit rotation matrices after applying the altitude-azimuth interpolation.  Figure \ref{sincos_cumulative_error_distribution} shows each Mueller matrix estimate in a different color.  The trigonometric function fitting results in an uncertainty of less than 0.1 between the estimates and the rotation matrix fits.  

\begin{figure*} [!h, !t, !b]
\begin{center}
\includegraphics[width=0.36\linewidth, angle=90]{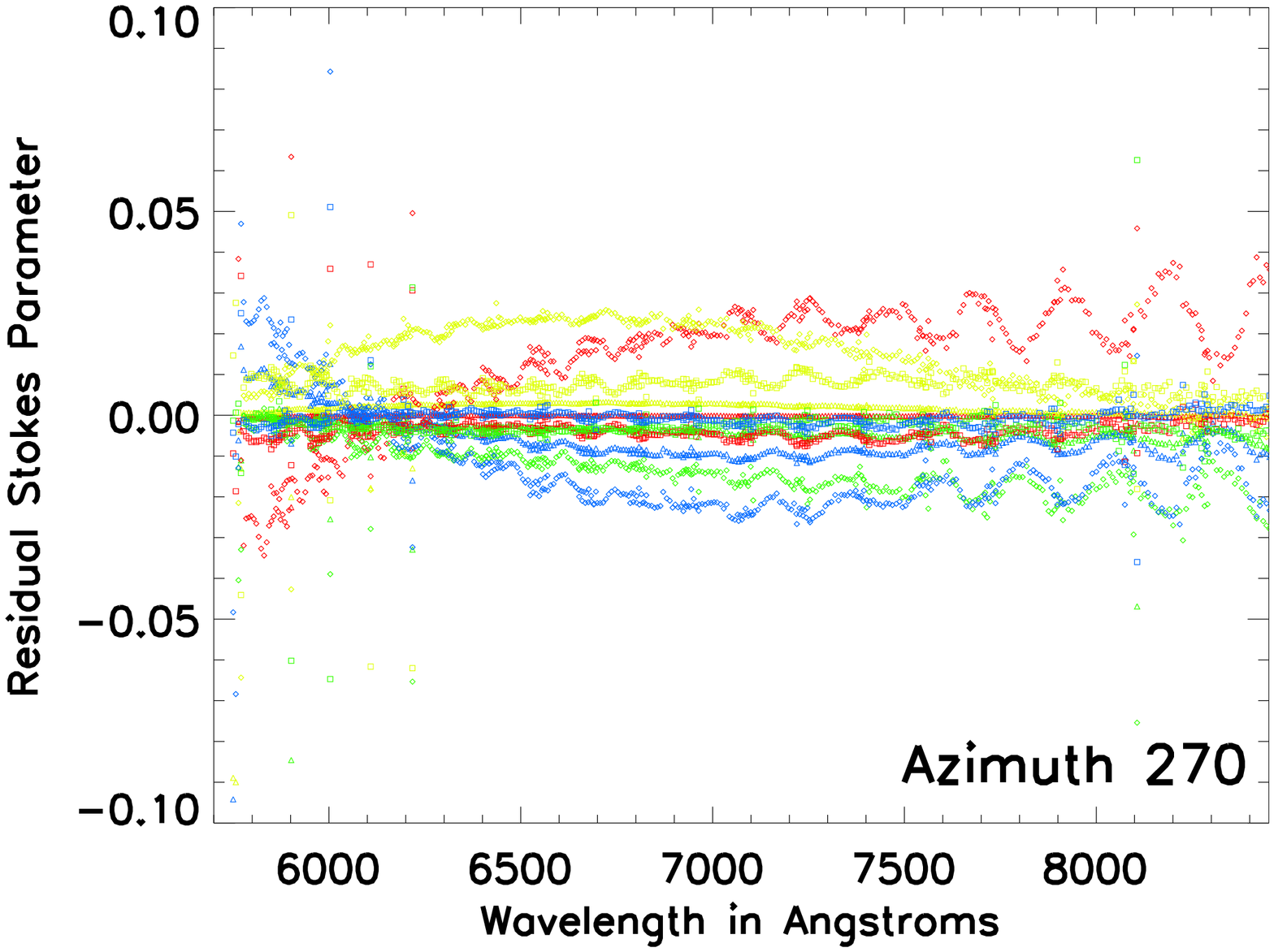}
\includegraphics[width=0.36\linewidth, angle=90]{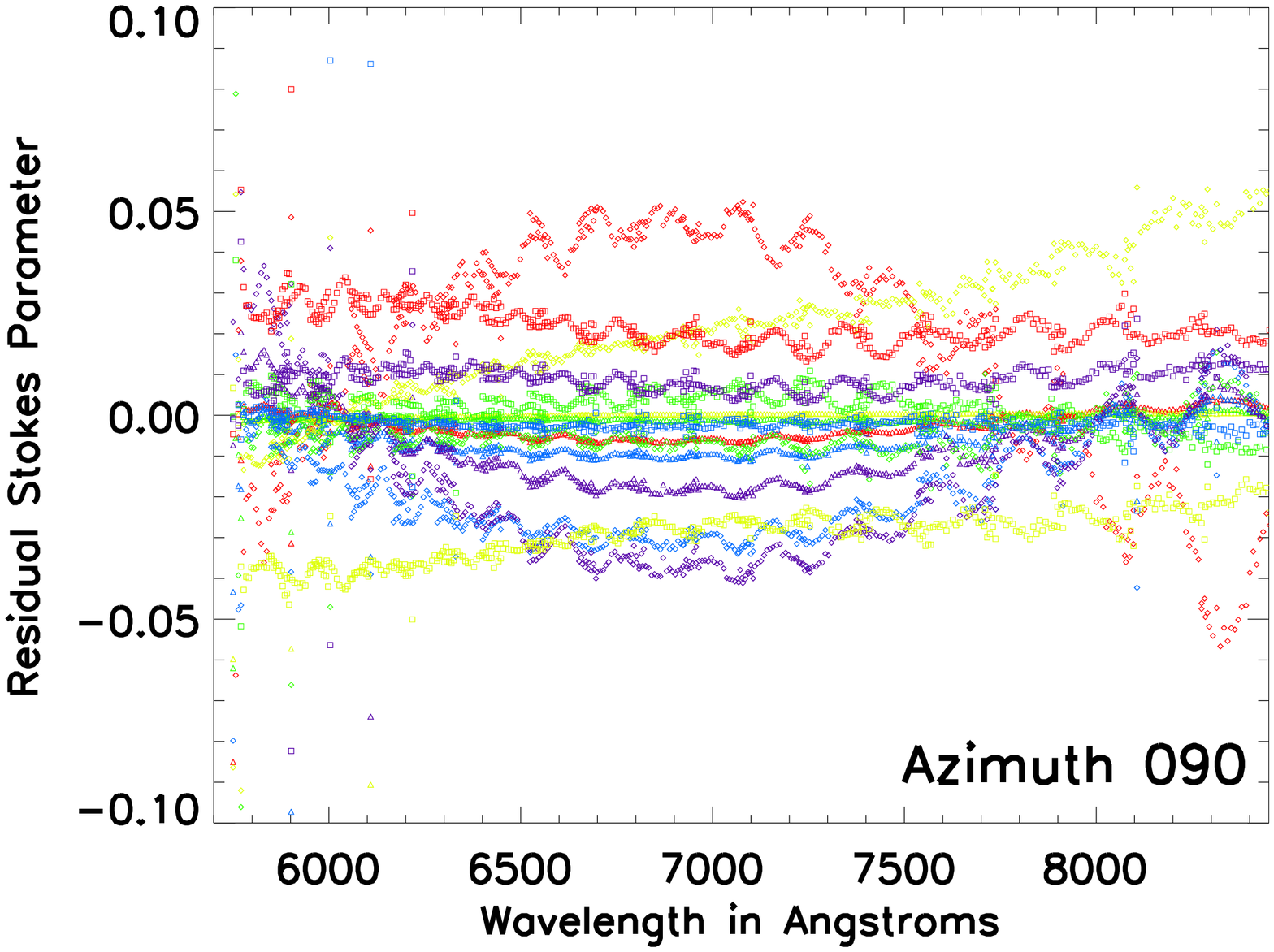}
\caption{ \label{residual_stokes_parameters_apr21} This Figure shows the difference between the demodulated and calibrated Stokes parameters and the predicted Rayleigh sky polarization spectra for April 21$^{st}$ 2012. Each panel shows observations calibrated at a different telescope azimuth. There are 4 to 5 independent observations at each telescope pointing.  Different colors represent residual Stokes parameters for individual observations computed as data minus model. The residuals are always below 0.1 across the wavelength range. Note - only a single calibration point per spectral order has been applied and the data is shown binned to 50 spectral samples per order. Note that these residual Stokes parameters give rise to the few degree rotational error between a measured Stokes vector and a model Stokes vector projected on the Poincar'{e} sphere as shown above in Figure \ref{derotated_data_histogram}.}
\end{center}
\end{figure*}

	One of the major sources of calibration uncertainty in the present approach is the stability of the liquid crystals in temperature and time. The individual measurements can calibrated and differenced from the theoretical daytime sky polarization to demonstrate the demodulation and cross-talk calibration errors. Figure \ref{residual_stokes_parameters_apr21} shows the difference between the theoretical Rayleigh sky polarization spectra and demodulated calibrated HiVIS measurements. In this Figure, April 21$^{st}$ 2012 data was used to derive Euler angles. This data was then polarization calibrated and subtracted from the theoretical Rayleigh sky polarization. The residual errors are typically less than 0.05 in any individual Stokes parameter. These residual errors were used to derive a rotational measure of absolute calibration accuracy and are the basis for the histograms of the angle between the measured calibrated Stokes vector and the theoretical daytime sky Stokes vector shown above in Figure \ref{derotated_data_histogram}.

\begin{figure} [!h, !t, !b]
\begin{center}
\includegraphics[width=0.89\linewidth, angle=90]{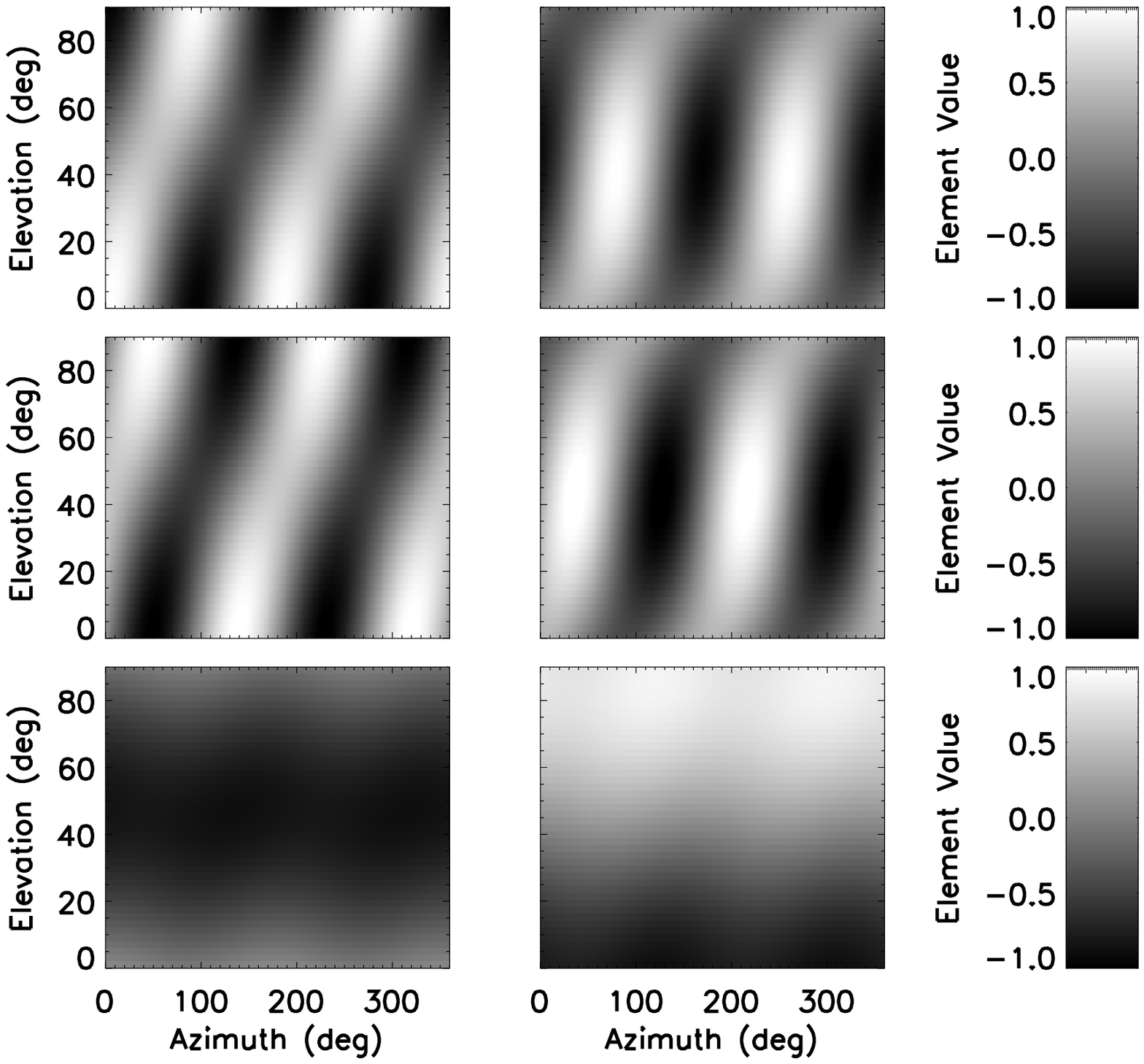}
\includegraphics[width=0.89\linewidth, angle=90]{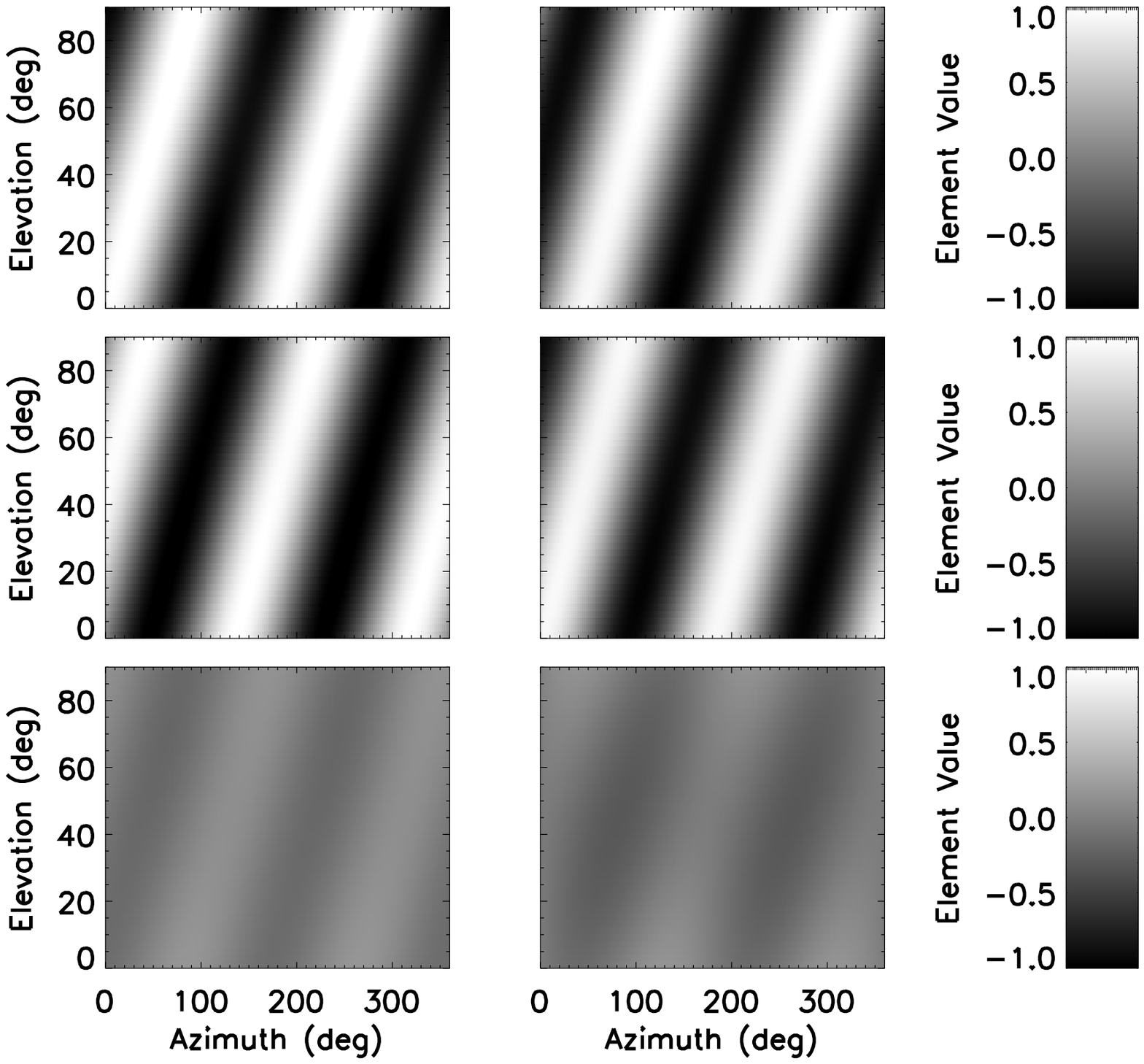}
\caption{ \label{sincos_with_wavelength} This Figure shows the six Mueller matrix estimates trigonometry function fitting for all altitudes and azimuths. The $QQ$ and $QU$ terms are on top. The $UQ$ and $UU$ terms are in the middle.  The $QV$ and $UV$ terms are on the bottom.  The top graphic shows spectral order at 6225{\AA}. The bottom graphic shows spectral order at 8601{\AA} with much reduced circular polarization cross-talk (low $QV$ and $UV$ values). The sinusoidal variation in the linear polarization terms represent the transformation between the altitude-azimuth frame and a Zenith based polarization frame.}
\end{center}
\end{figure}

	We find that the likely cause of the ripples seen at small wavelength scales within individual spectral orders is caused by slight changes in liquid crystal properties. In Figure 3 of \cite{Harrington:2010km}, we showed a transmission function using a very small beam footprint using a Varian spectrophotometer. This Varian is effectively a lab-calibrated commercial scanning monochrometer. This instrument also showed small-scale ripples in the overall transmission function that were highly dependent on placement of the liquid crystal. Similar ripples were seen in Figure 5 of \cite{Harrington:2010km} when the temperature of the device was changed. Liquid crystal stability will likely be a limitation of our system. We note that the calibrations applied here only use one telescope Mueller matrix element per spectral order.  Potentially performing calibrations with more data at higher spectral sampling could improve the calibration procedures.

	The wavelength dependence of the telescope Mueller matrix is very slow.  Figure \ref{sincos_with_wavelength} shows the six Mueller matrix estimates after being fit to simple trigonometric functions of altitude and azimuth.  The top panel shows the 6225{\AA} spectral order while the bottom panel shows the 8601{\AA} spectral order. The linear polarization terms are dominated by the transformation between an altitude-azimuth based frame and a Zenith frame.  The circular polarization terms are quite strong at 6225{\AA} but tend towards zero at 8601{\AA}.  Given this gradual wavelength dependence, we find that one calibration per spectral order is sufficient for this demonstration.

\subsection{Example Calibration of $\epsilon$ Aurigae}
	
	Epsilon Aurigae is a very bright target with a known strong and variable H$_\alpha$ polarization signature \citep{Harrington:2009dz, 1986ApJ...300L..11K, 2011SASS...30..103C, 1985eepa.rept...33K, 2010JSARA...4....2H, 2012AIPC.1429..140H, 2012JSARA...5...51B, Stencel:2013td}. We have a long and detailed set of $\epsilon$ Aur observations taken with ESPaDOnS at CFHT for comparison. We have never detected any circular polarization in H$_\alpha$ in the 2006 to 2012 time frame. As an example of this stars H$_\alpha$ line profile, we compiled 42 independent measurements in to blocks where the target shows substantially different polarization and intensity line profiles.  Figure \ref{epsaur_specpol_espadons} shows the observations broken in to 9 main observing blocks covering summer and winter observing seasons over the 6 years. We followed our recipe for post-processing of the Libre-Esprit reduced ESPaDONs spectra outlined in the appendix of \citet{Harrington:2009ka}. Essentially our process combines the reduced spectra from individual spectral orders and averages pixels to a more uniform spectral sampling.  In the case of Figure \ref{epsaur_specpol_espadons} the effective spectral sampling is at about 35pm per pixel for a 1-point sampling of R$\sim$170,000 data. For the purposes of this calibration we note that the polarized H$_\alpha$ profiles have amplitudes of 0.5\% up to 1\% while Stokes $v$ signatures are undetectable to within the signal-to-noise thresholds. The effective signal to noise ratios in both the ESPaDOnS data sets is over 1,000 per pixel and the relevant spectral features are very broad. Our HiVIS charge-shuffled spectra are sampled at $\sim$44pm per pixel at H$_\alpha$ and we highly over-sample the delivered optical resolution of 12,000 to achieve high signal-to-noise ratios efficiently on bright targets.

\begin{figure*} [!h, !t, !b]
\begin{center}
\includegraphics[width=0.70\linewidth, angle=90]{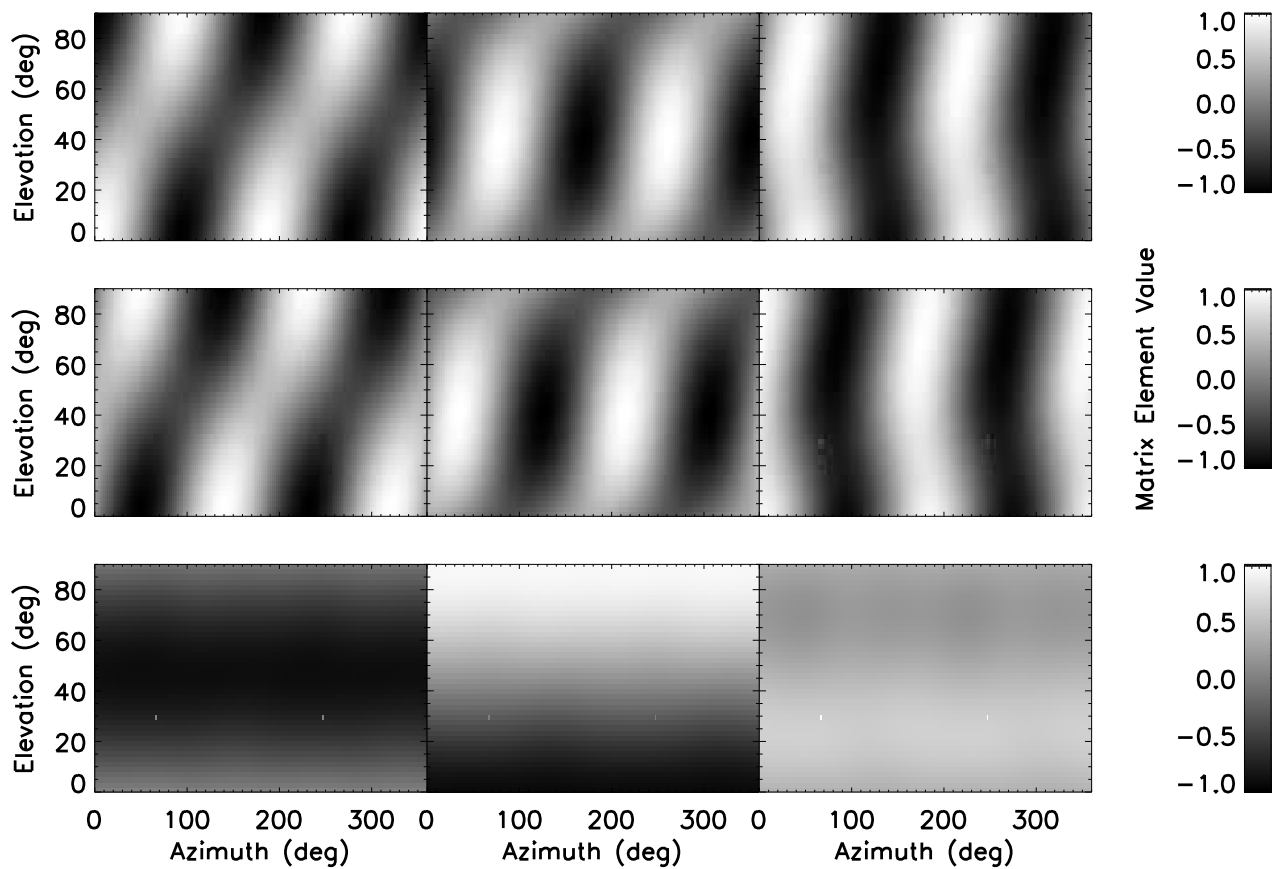}
\caption{ \label{mueller_matrix_altazi} An altitude-azimuth map of the system Mueller matrix cross-talk elements for the 6560{\AA} spectral order.  The best fit rotation matrix was fit to the Mueller matrix estimates elements following our standard procedure.  The Mueller matrix element az-el maps are all linearly scaled from black at -1 to white at +1. Each element is plotted from the horizon to the Zenith for the full 360$^\circ$ azimuth range.}
\end{center}
\end{figure*}

	During October 23, 2013, a large set of observations were recorded on the star $\epsilon$ Aurigae while performing our daytime sky calibration mapping procedures.  We collected 30 individual polarization measurements over the course of one night.  Over the observations, azimuths ranged from 055$^\circ$ through transit to 330$^\circ$. Elevations were between 34$^\circ$ and 67$^\circ$. Each full-Stokes $\epsilon$ Aur observation set was reduced and demodulated using the standard observing sequence with our polarization calibration unit. The signal-to-noise ratio for each individual $quv$ measurement was between 650 and 900 for all 30 independent $quv$ measurements.

\begin{figure} [!h, !t, !b]
\begin{center}
\includegraphics[width=0.75\linewidth, angle=90]{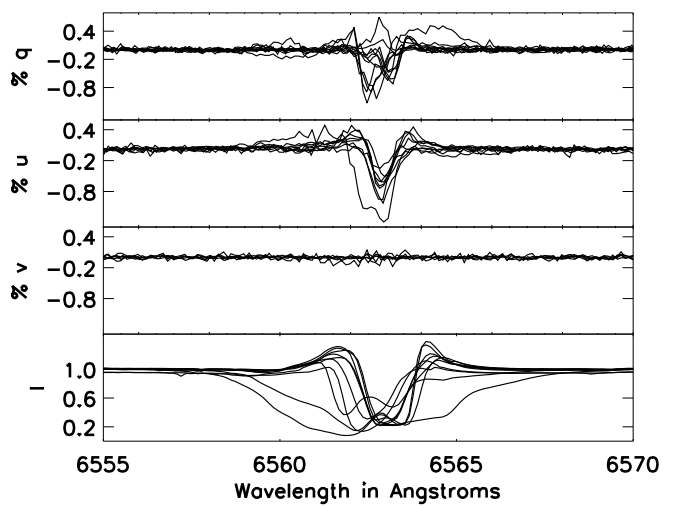}
\vspace{-4mm}
\caption{ \label{epsaur_specpol_espadons} This Figure shows the $\epsilon$ Aurigae H$_\alpha$ line profiles observed with ESPaDOnS between 2006 and 2012. This star has shown strong variable $qu$ profiles but Stokes $v$ was always zero within the statistical noise. See text for details.}
\end{center}
\end{figure}

\begin{figure} [!h, !t, !b]
\begin{center}
\includegraphics[width=0.75\linewidth, angle=90]{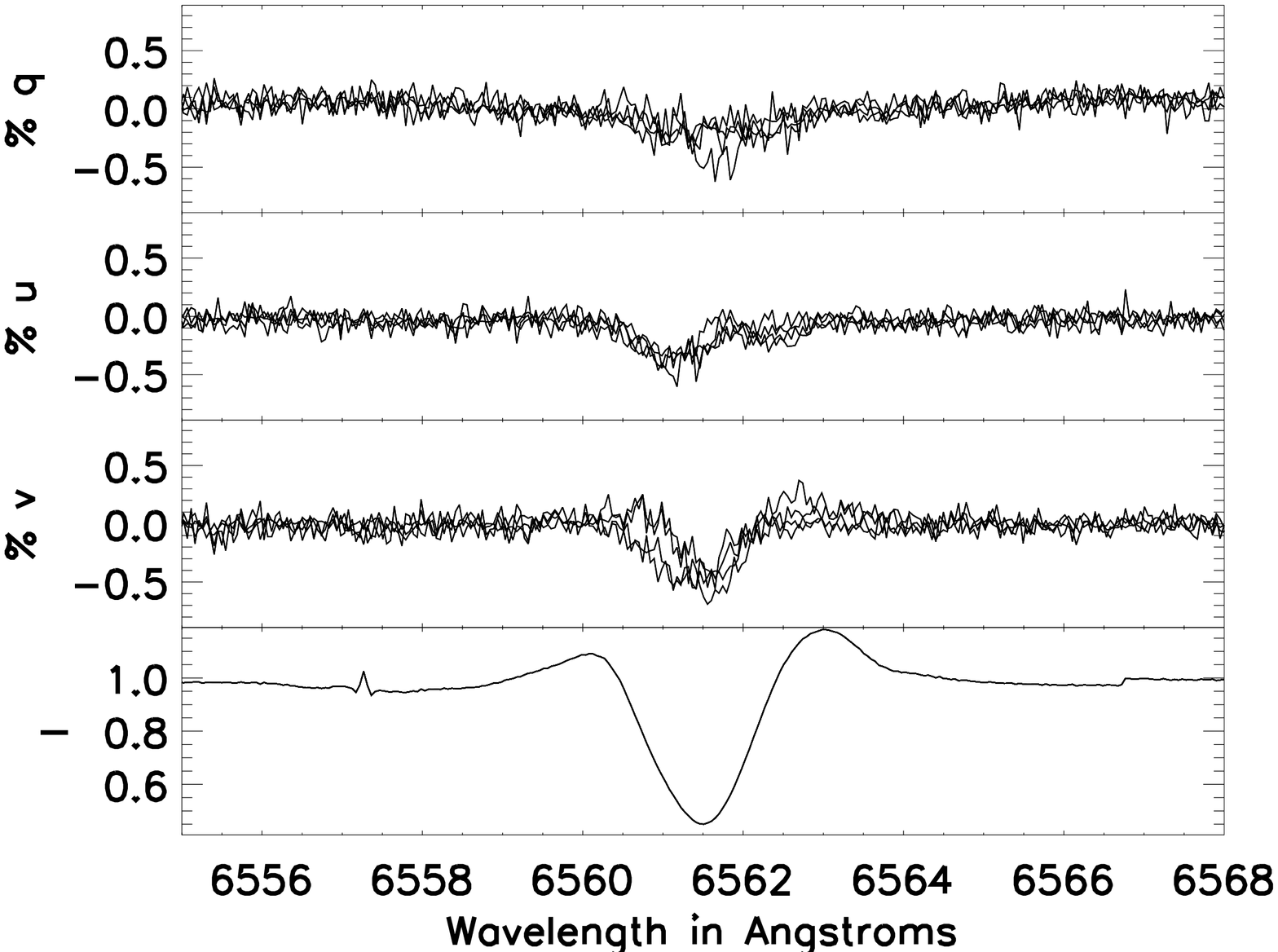}
\vspace{-2mm}
\caption{ \label{epsaur_specpol_uncalibrated} This Figure shows the average $quv$ data sets in 4 main telescope pointing blocks for $\epsilon$ Aurigae observed with HiVIS on October 23$^{rd}$ 2013. No telescope Mueller matrix calibration has been applied. The input signal is $qu$ as shown by ESPaDOnS observations but the detected HiVIS signature is much larger in $v$ due to the telescope cross-talk. The 4 independent $quv$ spectra show some spectral variation due to changes in the telescope cross-talk as the optics change with time. See text for details.}
\end{center}
\end{figure}

	There were 4 separate clusters of data recorded at telescope pointings of (051,35), (045,52), (011,67) and (333,64). There are 5, 10, 5 and 10 consecutive exposures in each of the associated observing blocks. Figure \ref{epsaur_specpol_uncalibrated} shows the $quv$ spectra averaged during these 4 observing blocks. The ESPaDOnS spectra show that the H$_\alpha$ signature is always contained entirely in linear polarization $qu$. However, since the linear to circular cross-talk is 100\% at some telescope pointings for HiVIS, our uncalibrated data shows this H$_\alpha$ signature largely in Stokes $v$. Some changes can be seen in the details of the $quv$ profiles, showing the changing cross-talk as the telescope tracks the star. 

\begin{figure} [!h, !t, !b]
\begin{center}
\includegraphics[width=0.75\linewidth, angle=90]{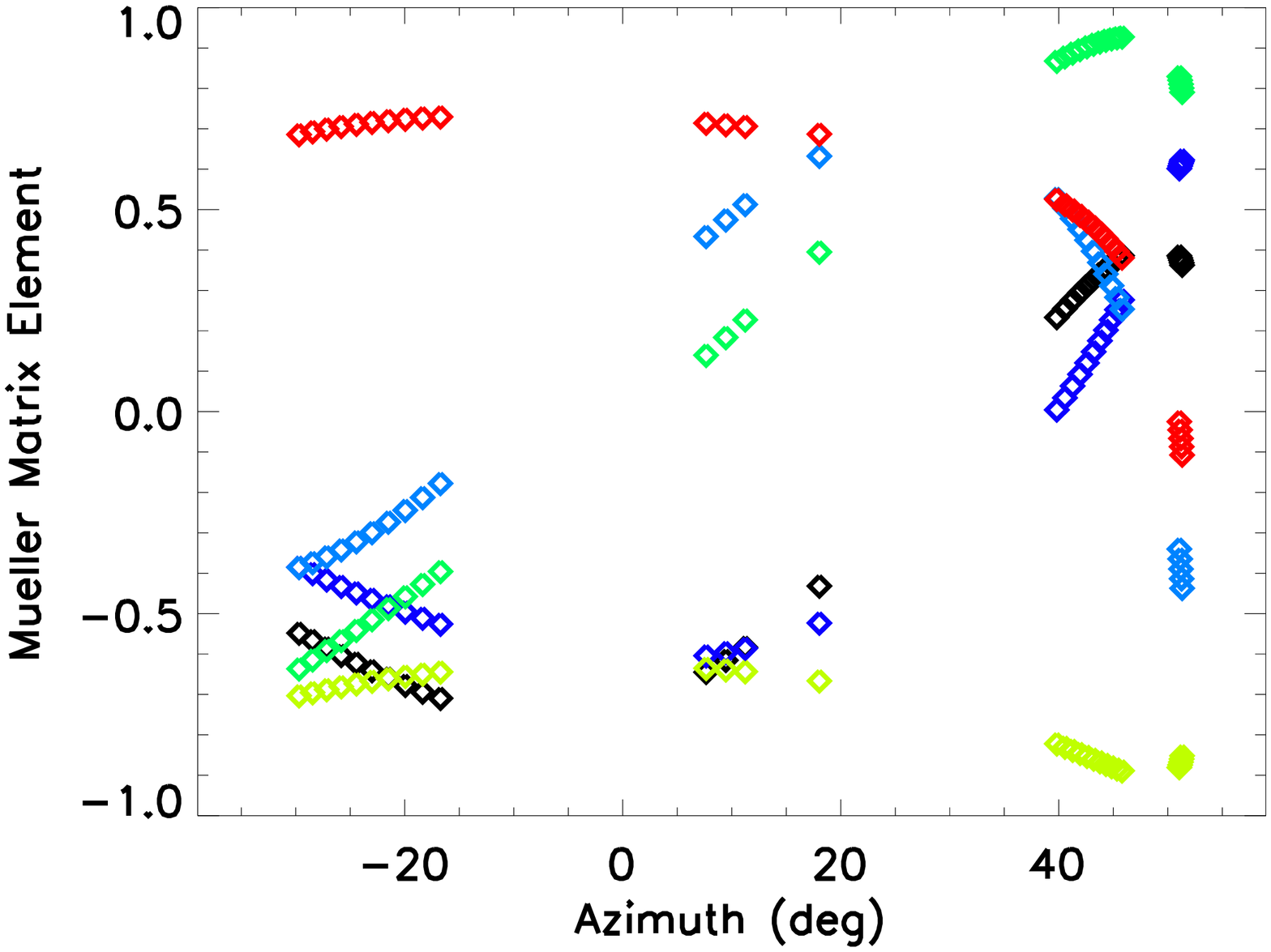}
\caption{ \label{rotation_matrix_elements_epsaur} The estimated system Mueller matrix elements changing with time for the 30 $\epsilon$ Aurigae data sets collected with HiVIS on the 23rd of October, 2013. This shows the time dependence of the system cross-talk elements of the Mueller Matrix varying quite substantially with telescope pointing and time. See text for details.}
\end{center}
\end{figure}

	Using the telescope Mueller matrix element estimates, the best-fit rotation-matrix was computed for the telescope pointing at each $\epsilon$ Aurigae data set following our procedures \citep{Harrington:2011fz}. The telescope Mueller matrix maps of Figure \ref{mueller_matrix_altazi} can be interpolated to the telescope pointing of every individual observation. Figure \ref{rotation_matrix_elements_epsaur} shows the telescope Mueller matrix estimates for every individual observation of $\epsilon$ Aurigae. The 4 observing blocks can be seen as clusters of points with similar Mueller matrix elements. For example, the far right side of Figure \ref{rotation_matrix_elements_epsaur} shows a cluster of 5 observations while $\epsilon$ Aurigae was rising at (azimuth, elevation) of (051,35).  The telescope pointing change during this time was largely elevation. On the far left side of Figure \ref{rotation_matrix_elements_epsaur}, a sequence of 10 observations is seen near an (azimuth,elevation) of (333,64) while the predominant telescope pointing change was in azimuth.

\begin{figure} [!h, !t, !b]
\begin{center}
\includegraphics[width=0.75\linewidth, angle=90]{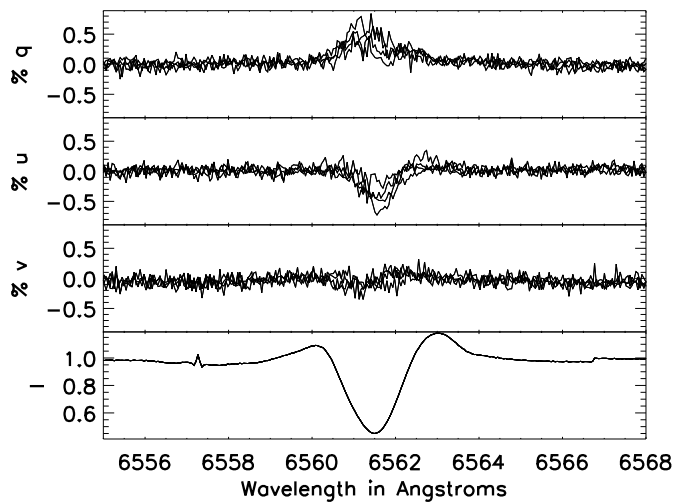}
\vspace{-2mm}
\caption{ \label{epsaur_specpol} This Figure shows the average of calibrated $quv$ data sets or Epsilon Aurigae observed with HiVIS on October 23$^{rd}$ 2013.  The 4 main observing blocks at similar telescope pointings were averaged to achieve high SNR. The signal is largely $qu$ as expected from ESPaDOnS observations.  The Stokes $v$ measurements are nearly zero. Some residual Stokes $v$ is expected when using calibrations derived over a 6 month time frame.  See text for details.}
\end{center}
\end{figure}

	These telescope Mueller matrix calibrations were applied to each $\epsilon$ Aurigae data set. Figure \ref{epsaur_specpol} shows the average demodulated and calibrated data sets for the 4 major observing blocks. The top three panels show Stokes $q$, $u$ and $v$ respectively. The bottom panel shows the continuum-normalized intensity profile for H$_\alpha$. The uncalibrated measurements showed the observed signature was dominantly Stokes $v$ with substantial $q$ and $u$. After applying the telescope calibrations, the H$_\alpha$ line profile signature is almost entirely contained in Stokes $q$ and $u$.

	There are several sources of error in these demonstration calibrations that can easily be improved.  First, we have only used one calibration per spectral order which gives some mild wavelength dependent residual errors across a spectral order.  Secondly, the daytime sky observations were recorded over a 6 month time frame. This leads to several degrees rotational errors  as seen in the histogram of Figure \ref{derotated_data_histogram}. Note that a Stokes parameter error of 0.05 corresponds to a 3$^\circ$ misalignment while a 0.2 error corresponds to an 11$^\circ$ misalignment. Since we only apply one calibration per spectral order, for an effective reduction in wavelength sampling by a factor of 4000, we get residual errors wavelength from the liquid crystal ripples and general wavelength dependent behavior as seen in Figure \ref{residual_stokes_parameters_apr21}. The third error comes from interpolating the best-fit rotation matrix to the telescope pointings of every individual observation. The cumulative distribution functions of Figure \ref{sincos_cumulative_error_distribution} show that the Mueller matrix element fitting process gives 1$\sigma$ errors at element values of roughly 0.04 to 0.10 for a 68\% error level.  This corresponds to additional rotational error levels of 2$^\circ$ to 6$^\circ$. 
	
\begin{figure} [!h, !t, !b]
\begin{center}
\includegraphics[width=0.95\linewidth, angle=0]{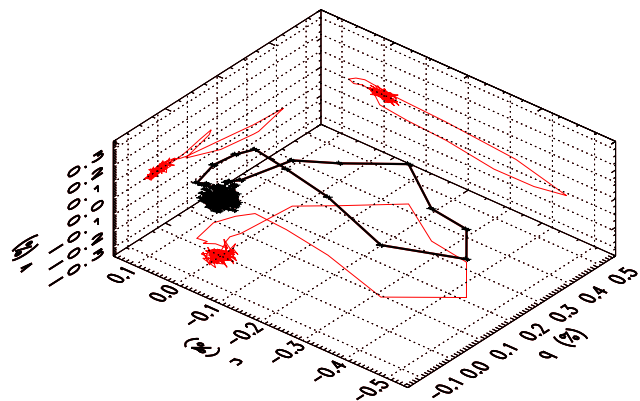}
\vspace{-2mm}
\caption{ \label{epsaur_quloop} This Figure shows the $quv$ behavior of calibrated $quv$ data sets or $\epsilon$ Aurigae observed with HiVIS on October 23$^{rd}$ 2013. The 4 main observing blocks at similar telescope pointings were combined after calibration to achieve high SNR of 5,000 to 7,600 in $quv$. The residual errors can be seen as a cluster of points around 0 with $\sigma<0.05\%$.  The signal is largely $qu$ with a linear polarization amplitude of 0.50\% as expected from ESPaDOnS observations.  The calibrated Stokes $v$ amplitude is roughly 0.14\%. This residual Stokes $v$ corresponds to a 15$^\circ$ rotational calibration error between linear and circular polarization. This error is expected when combining all the uncertainties in our calibration demonstration derived over a 6 month time frame.  See text for details.}
\end{center}
\end{figure}

	The calibrated $quv$ behavior for $\epsilon$ Aurigae is shown in Figure \ref{epsaur_quloop}. To show the $quv$ behavior at high signal to shot-noise ratios, we combined all individual calibrated spectra and then averaged 4 pixels in the spectral direction. The signal to noise ratio of these average spectra is between 5700 and 7700 as estimated from the standard deviation of the noise of the continuum adjacent to the H$_\alpha$ line.  After the 4-pixel averaging, we sample our R$\sim$12,000 spectra at an effective sampling of 34,000 for 3 points per resolution element.  The residual Stokes $v$ signature in Figure \ref{epsaur_specpol} is expected is an amplitude of 0.14\% while the $qu$ amplitude is at 0.50\%.  This corresponds to a misalignment of 15.6$^\circ$. As shown in Figure \ref{derotated_data_histogram}, we expect leakage of this magnitude considering all the instrument changes in the long time span between calibrations and observations.

\section{Conclusions}

	We have demonstrated here general techniques that overcome both calibration-type uncertainties (I) and measurement stability issues (II). SNRs above 0.01\% were achieved at full spectral resolution using calibration sources. On-sky calibrated observations of $\epsilon$ Aurigae achieved SNRs of 5000 to 7600 at R$=$34,000 sampling. Using our new daytime sky calibration technique, we have independently calibrated the polarization response of the instrument and telescope as functions of altitude and azimuth in a system with 100\% cross talk. In this altitude-azimuth telescope the polarization cross-talk calibration has a simple functional form on-sky and can be done easily on a routine basis to assure calibration stability. The cross-talk calibration was tested on a stellar target $\epsilon$ Aurigae with good results. Several sources of calibration algorithm uncertainty have been identified and future improvements are expected.

	The quality of our cross-talk calibration method was shown to be better than 4\% in an individual Stokes parameter if calibrations are taken within a day of the observations at the same telescope pointing. This corresponds to a 2$^\circ$ Poincare-sphere rotational uncertainty of the polarization orientation. Calibrations taken even a month after observations still provide 5$^\circ$ rotational accuracy which corresponds to about 10\% of an individual Stokes parameter. Our assumptions of a weakly depolarizing our telescope and instrument are reasonable given the degree of polarization measured for polarized standard stars as well as for the high measured daytime sky degree of polarization maximum ($\delta_{max}$).

	We were able to measure both polarized standard stars and unpolarized standard stars with smooth reproducible continuum polarization at the 0.1\% level even in the presence of a $>$2\% inter-order artifact. We applied firmware and software upgrades to accomplish charge shuffling with a CCD that was not originally designed for this purpose. Many of our instrument stability issues (type II) were overcome while improving our measurement efficiency. This new combination of charge shuffling, liquid crystal modulation and daytime sky calibration can deliver calibrated, high precision observations for use in a variety of stellar spectropolarimetric applications when high signal-to-noise data are obtained. 

	For other night-time instruments there are some relevant lessons learned from this study. Often, the major system limitations in high signal-to-noise applications is the ability to collect photo-electrons rapidly and suppress instrumental errors. We found that bidirectional CCD charge shuffling yielded significant calibration and efficiency gains. Multiple beams and storage buffers can be used without readout, reducing overheads. Modulation can be performed much faster than the exposure time. Operational flexibility is easily controlled via software. 
	
	Optical stability is also an important improvement. The benefits of improved guiding, reduced slit height, 1Hz modulation and having complete polarimetric information within a single exposure allowed sensitive continuum polarization measurements with the HiVIS spectropolarimeter. Fast (kHz and higher) modulation are not essential to substantially remove many sources of instrument instability and pointing jitter. These optical improvements will also remove sources of calibration uncertainty.

	The daytime sky calibration was readily adapted to this instrument. The telescope Mueller matrix elements de-rotated observations to a Poincare rotational accuracy of a few degrees with no night observing time lost. For AEOS, the altitude-azimuth dependence of the telescope Mueller matrix has a relatively simple functional form, which would be even simpler in equatorial or non-coud\'{e} optical systems. Calibrations at relatively few telescope pointings are sufficient to capture most of the observed Mueller matrix element variation. As  night-time and solar telescopes adopt polarimetric instruments in non-optimal optical configurations we believe the calibration strategies described here will find widespread applications.

\section{Acknowledgements}
Dr. Harrington would like to acknowledge support from the InnoPol grant: SAW-2011-KIS-7 from Leibniz Association, Germany. This work was supported by the European Research Council Advanced Grant HotMol (ERC-2011-AdG 291659). Dr. Kuhn acknowledges the NSF-AST DKIST/CryoNIRSP program. This program was partially supported by the Air Force Research Labs (AFRL), Boeing and the National Science Foundation. We'd like to thank Peter Onaka and John Tonry of the Pan-STARRS detector group at IfA for helping with replacement of one CCID20 array on the HiVIS focal plane mosaic. This work made use of the Dave Fanning and Markwardt IDL libraries.

\begin{appendix}
\section{Scattered Light as polarization contamination}

	The optical performance and stability of a spectrograph impacts our ability to calibrate systematic errors and to measure stable continuum polarization. Optical imperfections cause scattered light, wavelength instabilities and spectropolarimetric issues with amplitudes above 0.1\%. These are also particularly difficult for intra- and inter-order stability which effect absolute polarization calibration. In order to estimate these effects, we performed a thorough optical stability analysis and detector characterization. 

\begin{figure} [!h, !t, !b]
\begin{center}
\hbox{
\hspace{1.5em}
\includegraphics[width=0.75\linewidth, angle=90]{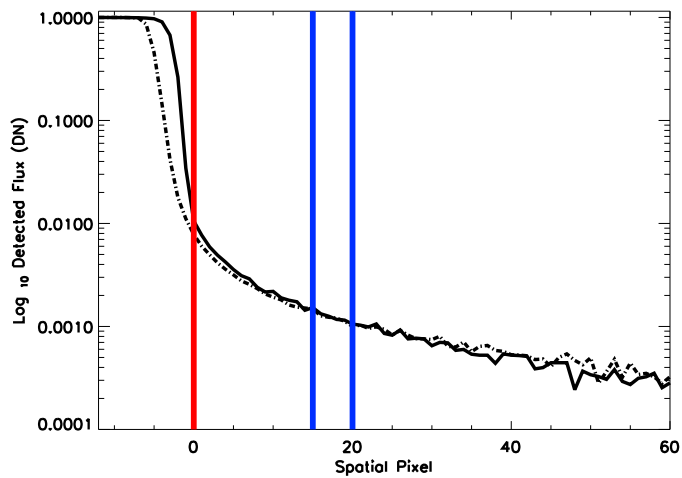}
}
\caption{ \label{lovis_psf_comp} This Figure shows the spatial profile delivered for the spectrograph in the LoVIS configuration with the flat-field lamp (April 26th 2011). There were no polarimetric components present in the beam. The spatial profile is shown on a log scale.  The red vertical line shows the edge of the extracted spectrum typically set in our data reduction pipeline. The blue lines show the region typically used to compute the background (dark, bias \& scattered light) extraction. }
\end{center}
\end{figure}

	Flat-field calibration images were computed and averaged to very high SNR to measure and remove the scattered light contamination. We chose to use the LoVIS configuration \citep{Harrington:2010km} with single spectral order imaged on the detector. Most of the CCD focal plane is not directly illuminated in this configuration allowing us to determine the wide-angle scattered light properties without inter-order contamination from adjacent spectral orders. Figure \ref{lovis_psf_comp} shows the spatial profile derived from a 100-exposure average of flat field measurements scaled to the peak flux. The dark and bias values have been removed from this stacked image with another 100-image combination of dark frames. This ensures high precision characterization of the scattered light in the spectrograph as the dark and bias are independently removed without choosing a specific region for dark subtraction. The red vertical line shows the edge of the spectral order as defined by the data reduction pipeline in the typical HiVIS charge shuffling mode. We use a 13 pixel half-width from the central spatial pixel for a total width of 27 pixels used in the spectral extraction. The flux levels at this order edge are typically below 1\% of the peak brightness. The blue lines show the region of the image typically used for background subtraction in the HiVIS data reduction pipeline. This region is usually another factor of 5 or more below the peak flux levels.  Similar results for scattered light are seen in the HiVIS configuration with both the achromatic rotating retarders and the liquid crystal variable retarders. 

	The efficient subtraction of scattered light is important for absolute polarization calibration. Any background intensity not effectively removed from all four polarized beams for each spectral order will depolarize and add other continuum polarization effects. There are 17 spectral orders extracted with HiVIS using the blue cross-disperser. There are two CCDs with two readout amplifiers per CCD. Each spectral order has pixels outside the illuminated region that are used for extracting an effective background. These backgrounds show scattered light at the few hundred count level.

\begin{figure} [!h, !t, !b]
\begin{center}
\includegraphics[width=0.75\linewidth, angle=90]{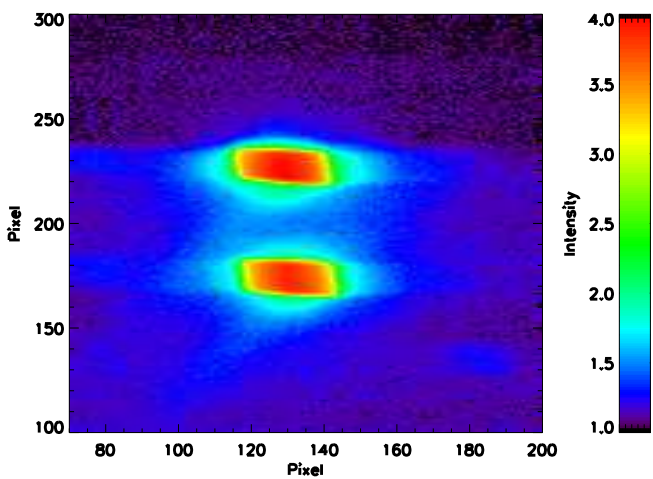}
\caption{ \label{ThAr2D_profile_july} This Figure shows a log-scaled extraction of Th-Ar arc line profiles computed using a stack of 40 individual exposures from July 25th 2012. A clear asymmetry is seen in the wings of the monochromatic slit-image at roughly 0.01\% of the peak intensity.}
\end{center}
\end{figure}

	Additional information about optical contamination in the presence of strong and spectrally narrow features is easily derived from arc lamp spectra. HiVIS has a Thorium-Argon (ThAr) lamp mounted on a calibration bench at the entrance to the coud\'{e} room. An example of a high-precision monochromatic slit image derived from ThAr exposures is shown in Figure \ref{ThAr2D_profile_july}. The red cross disperser was used in HiVIS with the 1.5$\arcsec$ slit. No charge shuffling was performed to illustrate the optical performance at high scattering angles. HiVIS has a spectral sampling of 4.5pm to 5.5pm per pixel corresponding to a 1-pixel sampling of R$\sim$150,000. The spectrum is highly over sampled. The Figure is log-scaled and one can easily see there are some asymmetries in the background flux in the spatial direction at the 0.1\% level. This is quite small and stable with minimal influence on the delivered polarimetric calibration accuracy. 

	This instrument has been used to measure spectropolarimetric profiles in highly distorted emission lines. Differential bleeding of intensity from different spectral locations can contaminate observations. Spectral lines from P-Cygni and Herbig Ae/Be stars can have intensity contrasts of $>$30 over very narrow spectral regions. The effective spectral resolution is typically derived by measuring the full-width-half-maximum of arc lines. However, at small signal levels a more complete characterization is necessary. To estimate potential leakage from different spectral orders, we computed the spectral profile for arc lamp lines using stacks of 40 images. These profiles show that we deliver $>$99\% concentration of monochromatic light within 30 spectral pixels. The differential contamination for HiVIS is quite small.

	In addition to spectral bleeding, any Savart plate can potentially cause degradation.  A Savart plate is manufactured so that the extraordinary and ordinary beams interchange between the two crystals and the two exit beams correspond to $e-o$ and $o-e$ pathways. However, imperfect construction can cause leakage in to the $ee$ and $oo$ beams  We had a special alignment procedure done at the manufacturer with a $<$10$^{-5}$ leakage specification. We also have done tests of our upgraded Savart plate and we find this beam leakage undetectable. This specification is critical to HiVIS as leakage would have directed two additional spectral beams in between polarized orders decreasing contrast and adding spectral leakage at $\sim$54 pixel separations.

\begin{figure} [!h, !t, !b]
\begin{center}
\hbox{
\hspace{1.5em}
\includegraphics[width=0.75\linewidth, angle=90]{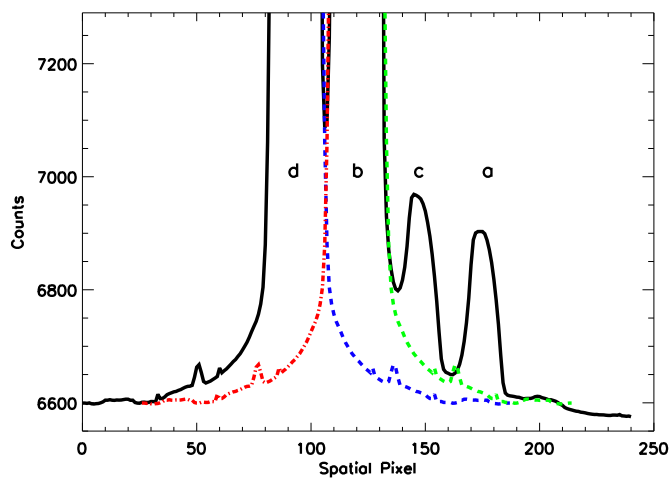}
}
\caption{ \label{100525_psf_bleeding} This Figure shows the computed spatial profile for the spectral order 10 with a wide dekker setting. This wide dekker exaggerates the intra-order contamination. There were no modulators in the beam. Charge shuffling of 27 pixels was done with polarization calibration unit polarizer in place. Beams $a$ and $b$ represent one set of orthogonally polarized spectra. Beams $c$ and $d$ represent another set of orthogonally polarized spectra. The asymmetry in intra-order contamination is easily seen where intense $d$ and $b$ beams strongly influence $c$ but not $a$. In double-differenced polarization calculations, this leads to spurious continuum polarization effects. }
\end{center}
\end{figure}

	The spatial width of the extracted spectra combined with the slit dekker width and any optical imperfections will cause intra-order contamination between modulated beams. This will create a systematic continuum polarization error when doing charge shuffling. The light that bleeds from one modulation state with polarization ($a-b/a+b$) in to the other modulation state with polarization ($c-d/c+d$) asymmetrically will cause continuum variations. As an illustration of this effect, Figure \ref{100525_psf_bleeding} shows the four beams for the 10$^{th}$ spectral order for HiVIS on May 25$^{th}$ 2010. In order to exaggerate the typical continuum effects, the slit dekker was set wider than nominal. The modulators (liquid crystals and rotating retarders) were removed to ensure that scattered light was the dominant error.

	One can see in Figure \ref{100525_psf_bleeding} that the asymmetric bleeding of one beam in to the other will cause different computed intensities even though no modulation or optical movement was present. The incomplete separation of beams lead to a systematic decrease in the detected degree of polarization for each of the HiVIS orders as well as an increase in the systematic error across each individual order.  The $b$ and $d$ beams had similar intensity and similar bleeding so these spectra were influenced little.  The $c$ spectra was much more substantially influenced because of it's low intensity (300 counts) and the proximity to the bright $b$ spectrum. The uneven disruption of the $a$ and $c$ spectra caused a much more substantial error in the $(c-d)/(c+d)$ polarization measurement.  

	In typical double difference calculations, the two methods do not give the same continuum polarization. The two methods in this nomenclature are:

\begin{equation}
q_{ab-cd}= \frac{Q}{I} = \frac{1}{2}(\frac{a-b}{a+b} - \frac{c-d}{c+d})
\end{equation}

and

\begin{equation}
q_{ac-bd}= \frac{Q}{I} = \frac{1}{2}(\frac{a-c}{a+c} - \frac{b-d}{b+d})
\end{equation}

\begin{figure} [!h, !t, !b]
\begin{center}
\hbox{
\hspace{0.8em}
\includegraphics[width=0.80\linewidth, angle=90]{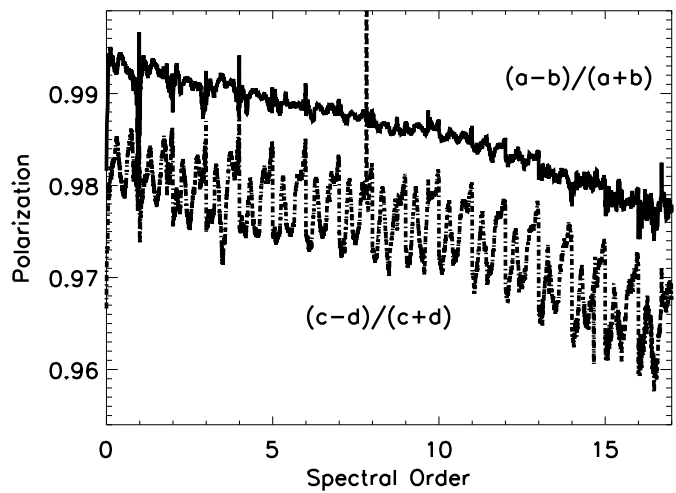}
}
\caption{ \label{100525_polarization_continuum} This Figure shows the derived Stokes $q$ continuum polarization with HiVIS using the polarization calibration unit and only the Savart plate analyzer. No modulators were present in the optical beam. The asymmetry between the (a-b) beams and (c-d) beams can be readily traced to the inter-beam contamination when strongly polarized signals are present.}
\end{center}
\end{figure}

	Without asymmetries in the optics, detector and modulation, these two equations should give identical results. In \cite{Harrington:2008jq} this is exactly what we find for HiVIS with only two well separated beams. The $q_{ab}$ calculation does a normalized ratio with spectra recorded simultaneously in time. The $q_{ac}$ calculation is a normalized ratio with spectra recorded on the same pixel for gain independence but with temporal changes between beams possible. If these four beams are not well separated and stable in time and completely unpolarized at the analyzer, continuum polarization errors will be present. As a demonstration, Figure \ref{100525_polarization_continuum} shows the computed polarization for this May 2010 experiment with the flat field calibration lamps. Since the modulators were removed and no optical motion occurred between modulation states, the derived Stokes $q$ polarization should be nearly 100\% and identical between the $ab$ and $cd$ calculations. Figure \ref{100525_polarization_continuum} shows that the $ab$ spectra have a very different contamination than the $cd$ spectra across individual spectral orders at the 0.5\% level.

	These asymmetries combined with typical optical instabilities (pointing, guiding, seeing, etc) conspire to add systematic continuum variations across spectral orders that are difficult to calibrate. This is a major limitation when stabilizing the continuum polarization of a slit spectropolarimeter.

\end{appendix}

\bibliographystyle{aa}
\bibliography{ms_arxiv}

\end{document}